\documentclass[aps,prx,twocolumn,groupedaddress,nofootinbib,superscriptaddress,showpacs]{revtex4-2}

\usepackage{hyperref}       % hyperlinks
\usepackage{url}            % simple URL typesetting
\usepackage{booktabs}       % professional-quality tables
\usepackage{amsfonts}       % blackboard math symbols
\usepackage{amssymb}
\usepackage{mathtools}
\usepackage{nicefrac}       % compact symbols for 1/2, etc.
\usepackage{microtype}      % microtypography
\usepackage{amsmath}
\usepackage{xspace}
\usepackage[T1]{fontenc}
\usepackage{dsfont}
\usepackage{pifont}
\usepackage[export]{adjustbox}
\usepackage{relsize}
\usepackage{xcolor,colortbl}
\usepackage{hhline}
\usepackage{makecell}
\usepackage{physics}
\usepackage{bbm}
\usepackage{changes}
\usepackage{tabularx}
\usepackage{amsthm}
\usepackage[nameinlink]{cleveref}
\usepackage{changes}
\usepackage{tikz}
\usepackage{color,soul}
\usepackage{multirow}
\usepackage{braket}

\newcommand\smallO{
  \mathchoice
    {{\scriptstyle\mathcal{O}}}% \displaystyle
    {{\scriptstyle\mathcal{O}}}% \textstyle
    {{\scriptscriptstyle\mathcal{O}}}% \scriptstyle
    {\scalebox{.7}{$\scriptscriptstyle\mathcal{O}$}}%\scriptscriptstyle
  }
  
\def\eg{\emph{e.g.}}
\def\ie{\emph{i.e.}}
\def\ii{\mathrm{i}}
\newcommand{\xmark}{\ding{55}}%

  \definecolor{masoncolor}{rgb}{0.98, 0.27, 0.62}

\begin{document}
\title{Complex networks with complex weights}
\date{\today}
%
%%%%%%%%%%%%%%%%%%%
%
\author{Lucas B\"ottcher}
\email{l.boettcher@fs.de}
\affiliation{Department of Computational Science and Philosophy, Frankfurt School of Finance and Management, 60322 Frankfurt am Main, Germany}
\author{Mason A. Porter}
\email{mason@math.ucla.edu}
\affiliation{Department of Mathematics, University of California, Los Angeles, CA, 90095, United States of America}
\affiliation{Santa Fe Institute, Santa Fe, NM, 87501, United States of America}
\date{\today}
\begin{abstract}
In many studies, it is common to use binary (\ie, unweighted) edges to examine networks of entities that are either adjacent or not adjacent. Researchers have generalized such binary networks to incorporate edge weights, which allow one to encode node--node interactions with heterogeneous intensities or frequencies (\eg, in transportation networks, supply chains, and social networks). Most such studies have considered real-valued weights, despite the fact that networks with complex weights arise in fields as diverse as quantum information, quantum chemistry, electrodynamics, rheology, and machine learning. 
Many of the standard network-science approaches in the study of classical systems rely on the real-valued nature of edge weights, so it is necessary to generalize them if one seeks to use them to analyze networks with complex edge weights. 
In this paper, we examine how standard network-analysis methods fail to capture structural features of networks with complex edge weights. We then generalize several network measures to the complex domain and show that random-walk centralities provide a useful approach to examine node importances in networks with complex weights.
\end{abstract}
\maketitle

\section{Introduction}
Network analysis has provided useful insights into many physical, biological, and social phenomena~\cite{newman2018networks}. There has been a wealth of research both about network structure and about
the effects of network structure on dynamical processes (including opinions and social influence, the spread of infectious diseases, and synchronization)~\cite{porter2016}).
If information about the intensity of interactions (and hence about the coupling strengths) between nodes is unavailable, unweighted networks provide a reasonable starting point to study structural features of systems in which one can describe nodes as either in contact (or otherwise interacting) or not in contact. However, in many applications, it is useful to use edge weights to account for the intensities or frequencies of interactions between pairs of nodes. For example, researchers have used weighted edges to describe the contact frequencies between individuals in human social networks~\cite{salathe2010high}, passenger flows in air-transportation networks~\cite{barrat2004architecture}, interactions between different parts of a protein molecule~\cite{ribeiro2014determination}, and much more.

The vast majority of research on weighted networks has focused on networks with real-valued edge weights~\cite{barrat2004architecture,newman2004analysis,onnela2005intensity,saramaki2007generalizations,horvath2011weighted} and node weights~\cite{li2022nodeweights}. 
However, networks with complex weights arise in many scientific and engineering applications (see Tab.~\ref{tab:examples}), and it is necessary to adapt and reformulate existing network-analysis methods to study them. In the present paper, we extend ideas from the analysis of network structure to networks with complex weights.

In quantum physics~\cite{heisenberg1925,schroedinger1926,dirac1930principles}, one defines wave functions and their unitary evolution in complex vector spaces. Recent experiments provide strong evidence that real-valued formulations of the standard framework of quantum physics~\cite{mckague2009simulating} fail to capture physical reality~\cite{renou2021quantum,chen2022ruling,li2022testing}. 
In classical physics (\eg, in fluid dynamics and in electrodynamics), complex values allow one to simplify mathematical descriptions of wave-like phenomena. 
Additionally, researchers in machine learning, neuroinformatics, and allied subjects have a long history of exploiting complex-valued weights in the analysis and application of artificial neural networks~\cite{DBLP:journals/corr/ReichertS13,hirose2012complex,lee2022complex}. Researchers have also examined the impact of complex weights in studies of neuronal dynamics~\cite{nemoto2002complex,gomez2016}.
\begin{table*}
\centering
\renewcommand*{\arraystretch}{2.0}
\begin{tabular}{llcc}\toprule
\multicolumn{1}{c}{\textbf{Subject}} & \multicolumn{1}{c}{\textbf{Some applications}} & \makebox[6em]{$W=W^\dagger$} & \multicolumn{1}{c}{\,\,\,\textbf{References}\,\,\,} \\ \hline
\,\,\,Quantum information  \,\,\,& \,\,\,  \makecell[l]{Description of quantum walks} \,\,\, & \,\,\, \checkmark \,\,\, & \,\,\, \cite{childs2002example,moore2002quantum,childs2004spatial,portugal2013quantum,boettcher2021classical,kubota2021quantum,kadian2021quantum,frigerio2021generalized} \,\,\, \\ \hline
\,\,\,Condensed-matter physics \,\,\,&\,\,\, \makecell[l]{Electron transport in quantum networks;\\interactions between electrons and magnetic fields} \,\,\,& \,\,\, \xmark\checkmark \,\,\, &\cite{wu1991quantum,liu1999electronic,Lieb2004,vasilopoulos2007aharonov}\\ \hline
\,\,\,Mathematical chemistry  \,\,\,& \,\,\,  \makecell[l]{Structural properties of molecules} \,\,\, & \,\,\, \xmark\checkmark  \,\,\, & \,\,\, \cite{lekishvili1997characterization,golbraikh2002novel,estrada2006atomic} \,\,\, \\ \hline
\,\,\,Electrodynamics  \,\,\,& \,\,\,  \makecell[l]{Impedance values and reflection/transmission\\ coefficients in electrical networks} \,\,\, & \,\,\, \xmark \,\,\, & \,\,\, \cite{paul2007analysis,strub2019modeling,alonso2017power,chen2017power,muranova2019notion,muranova2020eigenvalues,muranova2021effective,muranova2022effective,muranova2022networks} \,\,\, \\ \hline
\,\,\,Electrostatics  \,\,\,& \,\,\,  \makecell[l]{Complex permittivities and permeabilities$^*$} \,\,\, & \,\,\, \xmark \,\,\, & \,\,\, \cite{paul2007analysis} \,\,\, \\ \hline
\,\,\,Rheology  \,\,\,& \,\,\,  \makecell[l]{Complex storage and loss moduli \\ of viscoelastic materials$^*$} \,\,\, & \,\,\, \xmark \,\,\, & \,\,\, \cite{lakes2009viscoelastic} \,\,\, \\ \hline
\,\,\,Computational social science  \,\,\,& \,\,\,  \makecell[l]{Social-network analysis} \,\,\, & \,\,\, \checkmark \,\,\, & \,\,\, \cite{hoser2005eigenspectral} \,\,\, \\ \hline
\,\,\,Machine learning  \,\,\,& \,\,\,  \makecell[l]{Complex-valued neural networks;\\graph neural networks;\\graph signal processing} \,\,\, & \,\,\, \xmark\checkmark \,\,\, & \,\,\, \cite{noest1987,noest1988discrete,noest1988associative,leung1991complex,benvenuto1992complex,kobayashi2010exceptional,kobayashi2016symmetric,zhang2021optical,spall2022hybrid,DBLP:conf/nips/ZhangHBPH21,he2022msgnn,DBLP:conf/pkdd/FurutaniSAHA19,DBLP:conf/aistats/CucuringuL0Z20} \,\,\, \\ \hline
\,\,\,Linear algebra  \,\,\,& \,\,\,  \makecell[l]{Study of Hermitian directed graphs} \,\,\, & \,\,\, \checkmark \,\,\, & \,\,\, \cite{liu2015hermitian,guo2017hermitian,mohar2020new,kubota2021periodicity} \,\,\, \\\bottomrule
\end{tabular}
\vspace{1mm}
\caption{\textbf{Summary of a variety of application areas of networks with complex weights.} For the indicated applications in electrostatics and rheology (which we mark with $^*$), one can use complex node weights to describe heterogeneous materials that are subject to heterogeneous electromagnetic and force fields. All other listed applications primarily use edge weights. A~\checkmark~indicates that a subject includes examples with Hermitian weight matrices, and a~\xmark~indicates that a subject includes examples with non-Hermitian weight matrices.}
\label{tab:examples}
\end{table*}

Network analysis has yielded insights both into individual quantum systems and in the study of quantum networks that one can construct using entangled states or physically interconnected subsystems~\cite{biamonte2019complex}. In quantum physics, one can think of Hermitian matrices that are associated with the Hamiltonian of an isolated quantum system as an adjacency matrix with real-valued diagonal terms (\ie, energy terms) and complex-valued off-diagonal terms, which describe changes in amplitude during a transition from one state to another. The connection between modularity maximization and node-occupation properties of random walks has been used to identify communities in such networks~\cite{faccin2014community}. Complex weights also arise in Dirac equations on networks~\cite{bianconi2021topological,bianconi2022dirac}, in quantum cellular automata~\cite{hillberry2021}, and in ``Vdovichenko's method''~\cite{vdovichenko65} to derive random-walk--based solutions of the Ising model on a two-dimensional (2D) lattice~\cite{morita1986justification,morita1990justification}.

In electrodynamics, one can interpret a network with real-valued edge weights as a network of resistors with weights that encode resistance values. One can use complex weights to describe impedance and admittance values in more general networks, such as in lumped-element models of coupled transmission lines~\cite{alonso2017power,chen2017power,muranova2019notion,muranova2020eigenvalues,muranova2021effective,muranova2022effective,muranova2022networks,strub2019modeling}, that include resistors, coils, and capacitors.

In applications of machine learning, allowing edge weights to take complex values can substantially improve the performance of artificial neural networks. In comparison to their real-valued counterparts, complex-valued neural networks can have better accuracies, convergence properties, and capacities to produce nonlinear decision boundaries (even for small numbers of neurons)~\cite{zhang2021optical}.\footnote{For example, one can use a single layer with two complex-valued neurons to represent conic sections (\ie, parabolas, circles, ellipses, and hyperbolas); a corresponding real-valued layer can only represent linear functions~\cite{zhang2021optical,spall2022hybrid}.} A variant of complex-valued neural networks called ``phasor neural networks'' have been used to construct associative memory~\cite{noest1987,noest1988discrete,noest1988associative}, and a complex-valued generalization of the original Hopfield network~\cite{amari1972learning,little1974existence,hopfield1982neural} has been trained in image-retrieval tasks using Hebbian learning~\cite{kobayashi2016symmetric}. To train complex-valued artificial neural networks with gradient-based methods, researchers developed a version of the backpropagation algorithm that can update complex-valued weights and biases~\cite{leung1991complex,benvenuto1992complex}.

In Tab.~\ref{tab:examples}, we indicate several application areas of networks with complex edge weights and/or complex node weights. This table does not give a rigid classification; instead, it summarizes a variety of areas in which complex adjacency matrices arise. Because of the connection between complex adjacency matrices and network descriptions of quantum transport of charged particles~\cite{avron1988adiabatic,Lieb2004}, some researchers also use the term ``magnetic adjacency matrix'' in this context~\cite{Smilansky2013,peron2020spacing,f2020characterization}. One can use magnetic adjacency matrices to construct ``magnetic Laplacian'', 
which have been integrated into graph neural networks to study node classification and edge inference (\ie, ``edge prediction'') in directed networks~\cite{DBLP:conf/nips/ZhangHBPH21,he2022msgnn}. 
{A very recent study~\cite{tian2023} examined the spectral properties of magnetic Laplacians that are} associated with networks with complex weights.
Many of the works that we list in Tab.~\ref{tab:examples} focus on Hermitian weight matrices $W$ (which satisfy $W=W^\dagger$). Complex weight matrices in electrodynamics, electrostatics, and materials science may not be Hermitian, as they describe heterogeneous materials or materials that are subject to heterogeneous fields.

In the present paper, we examine several issues with the application of standard network-analysis methods to networks with complex weights. We also examine a variety of connections to related physical systems that help us interpret the meaning of complex weight matrices (\ie, weighted adjacency matrices). In Sec.~\ref{sec:compl_weights}, we give a mathematical definition of networks with complex weights and we then discuss the relationship between such networks and random walks and opinion
consensus~\cite{masuda2017random}. 
As analogues of classical random walks and DeGroot consensus dynamics, we show that Hermitian complex weight matrices induce continuous-time quantum walks (CTQWs) and phase-synchronization dynamics that are related to the Schr\"odinger--Lohe model, which is a generalization of Kuramoto dynamics to non-Abelian oscillators and quantum oscillators~\cite{lohe2009non,lohe2010quantum,choi2014quantum,choi2016practical,antonelli2022}. In Sec.~\ref{sec:local_measures}, we identify another connection to phase synchronization by defining appropriate local network measures~\cite{barrat2004architecture}, such as a complex-valued node strength and a complex-valued weighted clustering coefficient. Importantly, although there exist connections between different dynamical systems and networks with complex weights, the evolution of a dynamical system on a network depends not only on an underlying weight matrix but also on the specific interaction rules that characterize that system~\cite{porter2016}. Nevertheless, as illustrated by several decades of research in network analysis, analyzing the mathematical properties (\eg, the spectra) of adjacency matrices and weight matrices can yield crucial insights into the stability and other properties of networked dynamical systems~\cite{newman2018networks,piet2015}.

In Sec.~\ref{sec:matrix_powers_walks}, we extend our discussion of walks on networks with complex weights and explain that one can interpret walks and their associated complex weights in terms of (1) interactions between charged particles and a magnetic field~\cite{Lieb2004} and (2) exchange statistics of indistinguishable particles~\cite{leinaas1977theory,wilczek1982magnetic,wilczek1982quantum,abelian_anyons2020,nakamura2020direct}. To mathematically characterize the structural differences between networks with binary, real, and complex weights, we study how different weight matrices affect graph energy (\ie, the sum of the absolute values of the eigenvalues of $W$) in Sec.~\ref{sec:graph_energy}. Graph energy is a common network measure in mathematical chemistry because of its connection to $\pi$-electron energy in tight-binding models~\cite{li2012graph}. 

One cannot directly apply certain network notions (such as eigenvector centrality and its generalizations) to networks with complex edge weights, as one must first have a matrix --- either a weight matrix or some other matrix, such as a function of a weight matrix --- that satisfies the Perron--Frobenius theorem~\cite{perron1907,frobenius1912}, for which we seek a real-valued and positive matrix to ensure that its leading eigenvector has strictly positive entries. In Sec.~\ref{sec:pf_eig_cen}, we discuss generalizations of the Perron--Frobenius theorem~\cite{rump2003perron,noutsos2012perron} and eigenvector centrality. 
Additionally, the inability to fully order complex numbers forces one to appropriately adapt other centrality measures if one desires to employ them. This issue becomes apparent for concepts like geodesic centrality measures, which are based on shortest paths. Approaches to compute such measures with Dijkstra's algorithm (when edge weights are positive) or the Bellman--Ford algorithm (when edge weights are either positive and negative) rely on comparing ordered quantities~\cite{bottcher2021computational,cormen2022introduction}. In Sec.~\ref{sec:bet_clo_cent}, we examine random-walk centrality measures that allow both real-valued and complex-valued edge weights~\cite{sole2016random,boettcher2021classical}. We thereby work with appropriate notions of centrality that allow us to avoid this issue.
Finally, in Sec.~\ref{sec:disc_concl}, we discuss our results and indicate some future directions in the study of networks with complex weights.

%%%%%%%

%
\section{Complex weights and their connection to random walks and opinion consensus}
\label{sec:compl_weights}
We consider networks in the form of weighted graphs $G=(V,E,w)$, where $V$ is a set of nodes, $E$ is a set of edges, and the function $w\colon E\rightarrow \mathbb{C}$ assigns a complex weight to each edge. The number of nodes is $N = |V|$. We use both an adjacency matrix $A\in\mathbb{R}^{N\times N}$ and a weight matrix (\ie, a weighted adjacency matrix) $W\in\mathbb{C}^{N\times N}$ to describe weighted edges between nodes. The entries $a_{ij}$ of the matrix $A$ are equal to $1$ if nodes $i$ and $j$ are adjacent and are equal to $0$ if they are not adjacent. Unless we state otherwise, we do not consider self-edges or self-weights. (That is, we take $a_{ii}=w_{ii}=0$.) To capture complex-valued relationships between nodes, we let the weight-matrix entries $w_{ij}=r_{ij}e^{\ii \varphi_{ij}}$ be complex numbers with magnitude $r_{ij}$ and phase $\varphi_{ij}$. If a network is undirected, then $r_{ij} = r_{ji}$ and $\varphi_{ij} = \varphi_{ji}$. If $a_{ij} = 0$, we set $w_{ij} = 0$.
In Fig.~\ref{fig:complex_weights}, we show examples of networks with real and complex weights. Both of the depicted networks are closed and directed triads. (A ``triad'' is a subgraph of a network that consists of three nodes.) 

%%%

\subsection{Weighted networks and linear diffusion dynamics}

\begin{figure}
    \centering
    \includegraphics[width=0.48\textwidth]
    {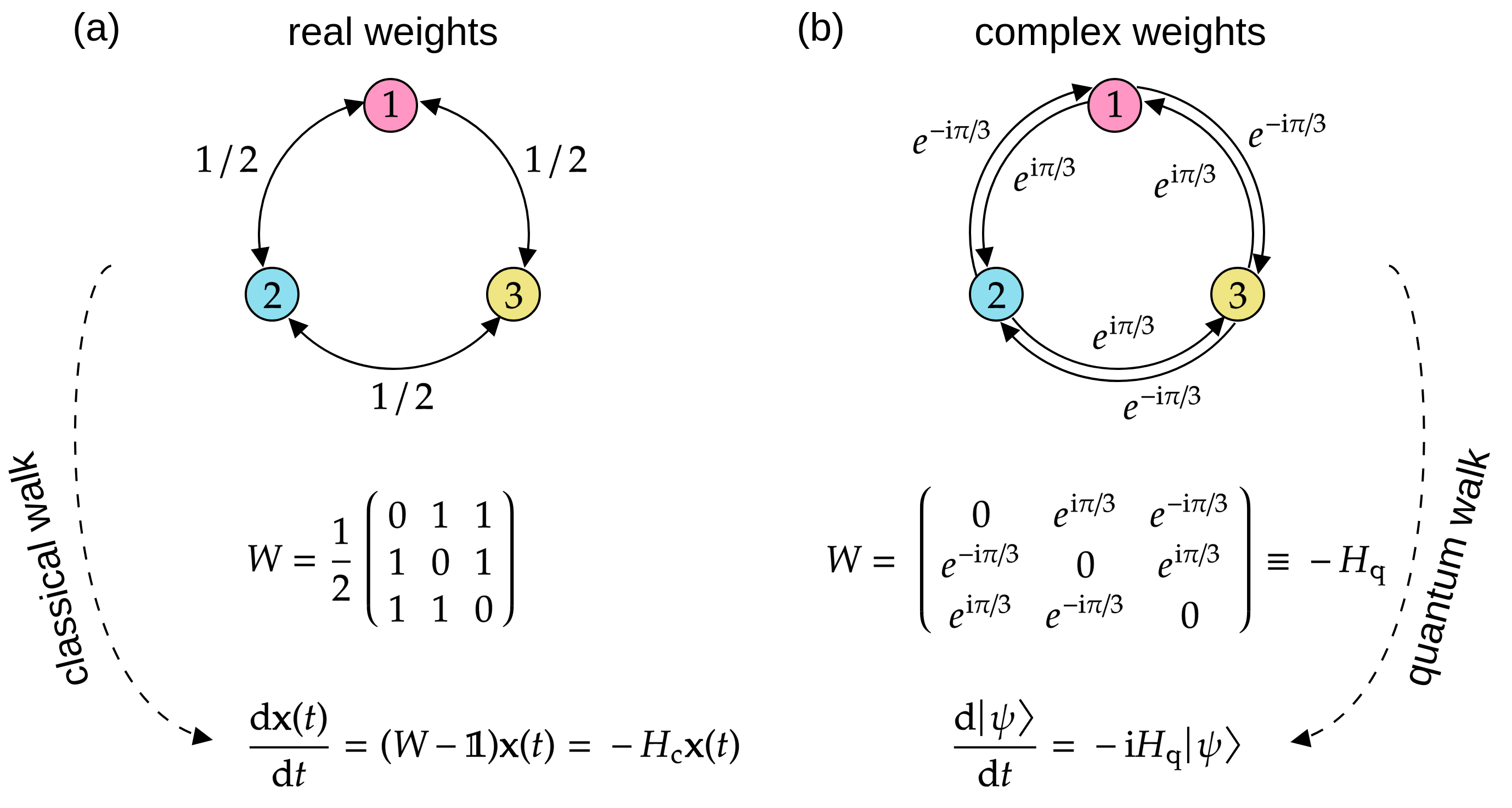}
    \caption{\textbf{Examples of networks with real and complex edge weights}. Each of these networks is a closed and directed triad.
    (a) The weight matrix $W$ is stochastic. It induces random-walk dynamics $\dot{\mathbf{x}}(t)=-H_{\rm c}\mathbf{x}(t)$, where $H_{\rm c}=\mathds{1}-W$ and $\mathbf{x}(t)$ is a probability vector whose entries $x_j(t)$ (with $j \in \{1,2,3\}$) give the probabilities of finding a random walker at each node $j$ at time $t$. (b) The weight matrix $W$ is Hermitian. It induces a continuous-time quantum walk $\dot{\ket{\psi}}=-\ii H_{\rm q} \ket{\psi}$, where $\ket{\psi}\in\mathbb{C}^3$ and $H_{\rm q}=-W$.
    }
    \label{fig:complex_weights}
\end{figure}

For a strongly connected network with nonnegative real weights $w_{ij}$ [see Fig.~\ref{fig:complex_weights}(a)], we can rescale $w_{ij}$ by mapping $w_{ij}\rightarrow w_{ij}/(\sum_{j=1}^N a_{ij} w_{ij})$ to obtain a stochastic weight matrix that induces the random-walk dynamics
\begin{equation}
    \frac{\mathrm{d}\mathbf{x}(t)}{\mathrm{d}t}=(W-\mathds{1})\,\mathbf{x}(t)=-H_{\rm c}\mathbf{x}(t)\,,
\label{eq:CTRW}
\end{equation}
where $H_{\rm c} = \mathds{1} - W$ is the classical random-walk Hamiltonian and $\mathbf{x}(t)$ is a probability vector with entries $x_j(t)$ (with $j \in \{1, \ldots, N\}$), which give the probabilities of finding a random walker at each node $j$ at time $t$. Under the aforementioned weight rescaling, one can thus always interpret a strongly connected network with nonnegative real weights in terms of transition probabilities of a classical random walker. We 
consider specific choices of $H_{\rm c}$ that are associated with left-stochastic and right-stochastic weight matrices.

For the left-stochastic weight matrix $W = A D^{-1}$, we obtain $H_{\rm c} = LD^{-1}$, where $L = D-A$ is the combinatorial graph Laplacian, $D = \mathrm{diag}(k_1,\ldots,k_N)$ is the degree matrix, and $k_i=\sum_j a_{ij}$ is the degree of node $i$ (\ie, the number of neighbors of node $i$). 
The stationary state of Eq.~\eqref{eq:CTRW} for $H_{\rm c} = LD^{-1}$ yields an occupation centrality measure that is proportional to node degree~\cite{faccin2013degree,boettcher2021classical}.

The random-walk evolution \eqref{eq:CTRW} is equivalent to the continuous-time DeGroot opinion-consensus model
\begin{equation}
    \frac{\mathrm{d}x_i(t)}{\mathrm{d}t}=\sum_{j=1}^N a_{ij}\left(x_j(t)-x_i(t)\right)
\label{eq:degroot}
\end{equation}
if we replace $H_{\rm c}  =LD^{-1}$ with $H_{\rm c} = L$~\cite{olfati2007consensus}. In the DeGroot model, we treat the underlying network as undirected (\ie, $a_{ij} = a_{ji}$) and $x_i(t)$ indicates the opinion of node $i$. 

For the right-stochastic weight matrix $W =D^{-1} A$, we obtain
\begin{equation}
    \frac{\mathrm{d}x_i(t)}{\mathrm{d}t}=\sum_{j=1}^N w_{ij}\left(x_j(t)-x_i(t)\right)
\label{eq:degroot_weighted}
\end{equation}
as a generalization of Eq.~\eqref{eq:degroot}. As in the original continuous-time DeGroot model \eqref{eq:degroot}, the nontrivial stationary state $\mathbf{x}^*$ of the weighted-network generalization 
\eqref{eq:degroot_weighted} is a consensus state, which entails that $x_i^* = x_j^*\equiv x^*$ for all nodes $i$ and $j$. Unlike in Eq.~\eqref{eq:degroot}, the quantity $\sum_i x_i(t)$ is, in general, not conserved in Eq.~\eqref{eq:degroot_weighted} because a stochastic weight matrix is symmetric only for regular graphs (\ie, for graphs in which each node has the same degree).

One can also establish a 
%similar 
connection between weighted networks and linear diffusion dynamics for networks with complex weights~\cite{childs2010relationship}. A Hermitian weight matrix $W$ (\ie, a weight matrix that satisfies $W = W^\dagger$) induces a CTQW that 
{evolves according to the Schr\"odinger equation}
\begin{equation}
   \ii \frac{\mathrm{d}\ket{\psi}}{\mathrm{d}t} = H_{\rm q} \ket{\psi}\,,
\label{eq:CTQW}
\end{equation}
where $\ket{\psi}\in\mathbb{C}^N$ and $H_{\rm q}=-W$ [see Fig.~\ref{fig:complex_weights}(b)]. The Hamiltonian $H_{\rm q}$ is the generator of time translation of a CTQW.

We use bra--ket notation. 
{In an $N$-dimensional Hilbert space (\eg, $\mathbb{C}^N$ equipped with the standard Hermitian inner product),} a ``ket'' $\ket{\psi}$ is a column vector and a ``bra'' $\bra{\psi}$ is the conjugate transpose of $\ket{\psi}$. The elements of the row vector $\bra{\psi}$ are thus complex conjugates of the corresponding elements of $\ket{\psi}$. As usual, $\braket{\psi|\phi}$ denotes the inner product that is associated with $\ket{\psi},\ket{\phi}\in\mathbb{C}^N$. 
The outer product of the two vectors is $\ket{\psi}\bra{\phi}$. In an $N$-dimensional vector space, the outer product is an $N \times N$ matrix.

The infinite-time mean
\begin{equation}
    \pi_{j} = \lim_{T\rightarrow\infty} \frac{1}{T} \int_0^T \braket{j|\rho(t)|j} \, \mathrm{d}t
    \label{eq:quant_occ}
\end{equation}
of a CTQW gives an occupation centrality measure for a network with complex edge weights~\cite{faccin2013degree}. In Eq.~\eqref{eq:quant_occ}, $\mathrm{d}t$ is an infinitesimal time step, $\rho(t)=\ket{\psi(t)}\bra{\psi(t)}$ is a density operator, and $\ket{j} \in \mathbb{C}^{N}$ is an orthonormal basis vector that satisfies
\begin{equation}
    \braket{i|j} = \delta_{i j}\,.
\end{equation}
%

%%%%

\subsection{Consensus dynamics, the Schr\"odinger--Lohe model, and synchronization}

We obtain a quantum-mechanical analogue of the DeGroot consensus dynamics \eqref{eq:degroot} by setting
\begin{equation}
    (H_{\rm q})_i = \ii \sum_{j=1}^N a_{ij} \left[\ket{\psi_j}\bra{\psi_i}-\ket{\psi_i}\bra{\psi_j}\right]\,,
\label{eq:SL_hamiltonian}
\end{equation}
which yields
\begin{equation}
    \ii \frac{\mathrm{d}\ket{\psi_i}}{\mathrm{d}t}=\ii \sum_{j=1}^N a_{ij} \left[\ket{\psi_j}-\braket{\psi_j|\psi_i}\ket{\psi_i}\right]\,.
\label{eq:quantum_consensus}
\end{equation}
Equation~\eqref{eq:quantum_consensus} preserves $\|\psi_i\|^2=\braket{\psi_i|\psi_i}$ for each state $\ket{\psi_i}$ (see App.~\ref{app:quantum_synch}) and synchronizes the relative phases between the states $\ket{\psi_i}$ and $\ket{\psi_j}$ (with $i \neq j$). It is {the} special case of the Schr\"odinger--Lohe model~\cite{lohe2009non,lohe2010quantum} with the Hamiltonian \eqref{eq:SL_hamiltonian}. 
Other versions of the Schr\"odinger--Lohe model include a Laplacian term in the Hamiltonian. These versions include both variants with an additional potential function \cite{choi2016practical} and variants without one \cite{choi2014quantum}.

If the states $\ket{\psi_i}$ satisfy $\|\psi_i\|^2 = 1$, then $\ket{\psi_i} = e^{-\ii \theta_i(t)}$. After the rescaling $a_{ij} \rightarrow a_{ij} K/(2N)$, the dynamical system~\eqref{eq:quantum_consensus} is equivalent to the Kuramoto model
\begin{equation}
    \frac{\mathrm{d}\theta_i(t)}{\mathrm{d}t}=\frac{K}{N}\sum_{j=1}^N a_{ij} \sin\left(\theta_j(t)-\theta_i(t)\right)
\label{eq:kuramoto}
\end{equation}
with a homogeneous coupling constant $K$ and a rotating reference frame in which all oscillators have the same natural frequency~\cite{kuramoto1975self}. The continuous-time DeGroot model \eqref{eq:degroot} corresponds to the linearization of Eq.~\eqref{eq:kuramoto} for small phase differences and $K = N$. {Lohe's generalization of Kuramoto dynamics to non-Abelian oscillators and quantum oscillators~\cite{lohe2009non,lohe2010quantum,choi2014quantum,choi2016practical,antonelli2022} is also related
to recent work on Kuramoto dynamics on high-dimensional spheres~\cite{lipton2021kuramoto}.}

In a network with a complex weight matrix $W$, we examine a weighted variant of the Hamiltonian~\eqref{eq:SL_hamiltonian}. This variant is 
\begin{equation}
    (H_{\rm q})_i = \ii \sum_{j=1}^N \left[w_{ji} \ket{\psi_j}\bra{\psi_i}-w_{ij}\ket{\psi_i}\bra{\psi_j}\right]\,.
\label{eq:weighted_Hamiltonian}
\end{equation}
The Hamiltonian $(H_{\rm q})_i$ in Eq.~\eqref{eq:weighted_Hamiltonian} is Hermitian if and only if $W$ is Hermitian (\ie, when $w_{ij}=\bar{w}_{ji}$ for all $i$ and $j$). Setting {$w_{ij} = K e^{-\ii \alpha}/(2N)$ and $w_{ji}=K e^{\ii \alpha}/(2N)$ yields the Sakaguchi--Kuramoto (SK) model\footnote{We use the term ``Sakaguchi--Kuramoto model'' because Ref.~\cite{sakaguchi1986soluble} lists Sakaguchi and Kuramoto as first and second authors, respectively. Some papers use the term ``Kuramoto--Sakaguchi model'' to refer to the same model.}
\begin{equation}
    \frac{\mathrm{d}\theta_i(t)}{\mathrm{d}t}=\frac{K}{N}\sum_{j=1}^N a_{ij}\sin\left(\theta_j(t)-\theta_i(t)+\alpha\right)\,,
    \label{eq:sakaguchi}
\end{equation}
where $\alpha$ is the phase lag and $|\alpha|\leq\pi/2$~\cite{sakaguchi1986soluble}. Like Eq.~\eqref{eq:kuramoto}, we can interpret Eq.~\eqref{eq:sakaguchi} as describing the evolution of coupled Sakaguchi--Kuramoto oscillators with the same natural frequencies in a rotating reference frame. Many papers have examined identical Kuramoto oscillators and identical Sakaguchi--Kuramoto oscillators in a rotating reference frame; see, \eg,~\cite{delabays2019kuramoto,ha2012basin,rodrigues2016kuramoto,ha2018emergence}. The empirical motivation of the phase-lag parameter $\alpha$ is that the common frequency of strongly coupled oscillators typically deviates from the mean of their natural frequencies. The SK model is relevant to various applications, such as the synchronization of coupled electrical oscillators~\cite{english2015experimental}. 

%%%%

\subsection{Quantifying classical and quantum consensus}

For classical consensus dynamics like \eqref{eq:degroot} and \eqref{eq:degroot_weighted}, we quantify the amount of consensus by calculating the order parameter
\begin{equation}
    r_{\rm c}(t)=1-\frac{1}{2 \Gamma}\|\mathbf{x}-\mathbf{x}^*\|_1\equiv 1-\frac{1}{2 \Gamma}\sum_{i=1}^N|x_i(t)-x_i^*|\,,
\label{eq:r_c}
\end{equation}
where $\Gamma=N x^*$ and $\mathbf{x}^*=(x^*,\ldots,x^*)^\top$ is the associated consensus state.

For the unweighted DeGroot model \eqref{eq:degroot} with $\sum_{i=1}^N x_i(0)=1$, we have $x^*=1/N$ (with $i \in \{1,\ldots,N\}$) and $\lim_{t\rightarrow\infty} r_{\rm c}(t)=1$. At time $t=0$, the consensus $r_{\rm c}(0)$ reaches a minimum $1/N$ for $x_i(0)=1$ and $x_j(0)=0$ (with $j\neq i$). At this minimum, the opinion of one node deviates maximally from the opinions of the other nodes. For the weighted DeGroot model \eqref{eq:degroot_weighted}, the stationary state is $\mathbf{x}^*=\lim_{t\rightarrow\infty}e^{(W-\mathds{1})t} \mathbf{x}(0)$.
\begin{figure}
    \centering
    \includegraphics{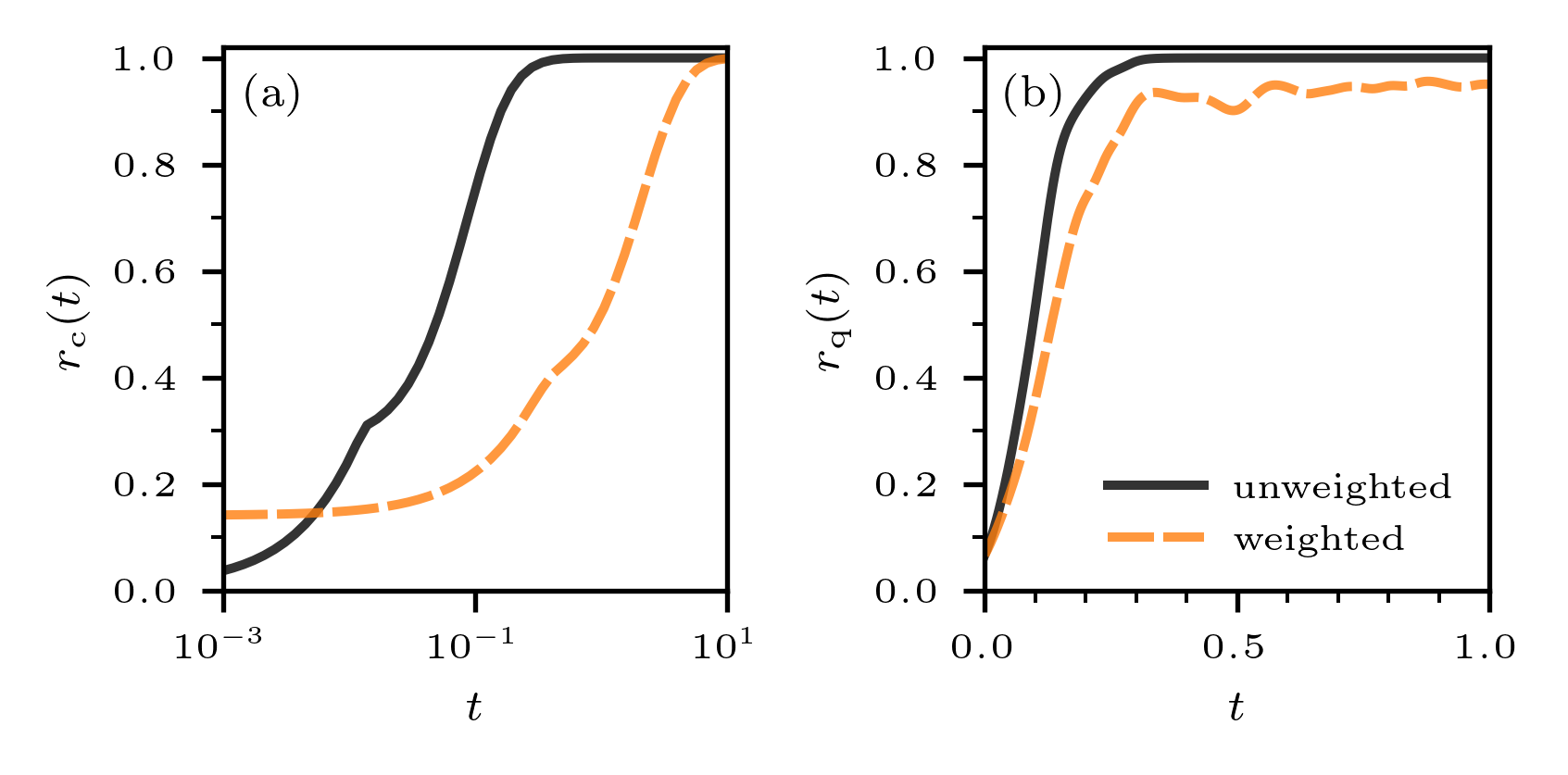}
    \caption{\textbf{Classical and quantum consensus.} The evolution of the (a) classical consensus \eqref{eq:r_c} and (b) quantum consensus \eqref{eq:r_q}. The underlying network is a $G(N,p)$ Erd\H{o}s--R\'{e}nyi (ER) network with $N=100$ nodes and connection probability $p=0.2$. The solid black curves indicate the {amount of consensus in the unweighted models}, and the dashed orange curves indicate the {amount of consensus in the weighted models.}
     For the weighted DeGroot model \eqref{eq:degroot_weighted}, we set $w_{ij} = a_{ij}/k_i$. Additionally, we set $w_{ij}=e^{-\ii \pi/4}$ and $w_{ji}=e^{\ii \pi/4}$ in the weighted Hamiltonian \eqref{eq:weighted_Hamiltonian}. In panel (a), we compute $\mathbf{x}(t)$ in $r_{\rm c}(t)$ by evaluating $e^{(W-\mathds{1})t} \mathbf{x}(0)$, where $x_1(0)=1$ and $x_j(0)=0$ for $j \in \{2, \ldots, N\}$. In panel (b), we use an implicit unitary integrator to solve Eq.~\eqref{eq:CTQW} with the Hamiltonian \eqref{eq:weighted_Hamiltonian}. We normalize each initial wave-function component $\ket{\psi_i(0)}$ (with $i \in \{1, \ldots, N\}$) to $1$, and we uniformly-randomly distribute their phases with mean $\pi/3$ and variance $25/3$.}
    \label{fig:consensus}
\end{figure}

For $\ket{\psi_i}=e^{-\ii \theta_i(t)}$, the corresponding order parameter for the quantum consensus dynamics \eqref{eq:quantum_consensus} is
\begin{align}
\begin{split}
    [r_{\rm q}(t)]^2&=1-\frac{1}{2 N^2}\sum_{i,j}\|\psi_i-\psi_j\|^2=\frac{1}{N^2} \sum_{i,j}e^{\ii\left(\theta_j(t)-\theta_i(t)\right)}\\
&=\frac{1}{N^2}\sum_{i,j}\cos\left(\theta_j(t)-\theta_i(t)\right)\,,
\end{split}
\label{eq:r_q}
\end{align}
so $r_{\rm q}(t)$ is equivalent to the order parameter of the Kuramoto model~\cite{kuramoto1975self}.

In Fig.~\ref{fig:consensus}, we show the evolution of $r_{\rm c}(t)$ and $r_{\rm q}(t)$ for classical and quantum consensus dynamics on a $G(N,p)$ Erd\H{o}s--R{\'e}nyi (ER) network with $N=100$ nodes and 
%connection\lb{Shall we maybe remove the word "connection"?} 
probability $p = 0.2$ for an edge to exist between two nodes. 
For the weighted DeGroot model \eqref{eq:degroot_weighted}, we set $w_{ij} = a_{ij}/k_i$. Additionally, we set $w_{ji} = e^{\ii \pi/4}$ and $w_{ij} = e^{-\ii \pi/4}$ in the weighted Hamiltonian \eqref{eq:weighted_Hamiltonian}. 
In Fig.~\ref{fig:consensus}, we observe that the unweighted consensus dynamics reach values $r_{\rm c}(t)$ and $r_{\rm q}(t)$ near $1$ faster than their weighted counterparts. 
We also observe that the weighted quantum consensus dynamics achieves smaller consensus values than their unweighted counterpart.

%%%%%

\section{Local network measures} 
\label{sec:local_measures}
Because the entries $w_{ij}$ of $W$ are complex-valued, the strength
\begin{equation}
    s_i = \sum_{j=1}^N w_{ij}
\end{equation}
of node $i$ is also a complex-valued quantity. In contrast to real-valued edge weights~\cite{barrat2004architecture}, $s_i$ does not provide a measure of importance or centrality of node $i$ because one cannot fully order complex numbers. 

To quantify the distribution of complex weights of the edges that are attached to node $i$, we define the normalized strength of node $i$ as
\begin{equation}
    \bar{s}_i = \frac{s_i}{k_i} = \frac{\sum_{j=1}^N w_{ij}}{\sum_{j=1}^N a_{ij}}\,.
\end{equation}
How do different values of the amplitudes $r_{ij}$ and phases $\varphi_{ij}$ affect the value of $\bar{s}_i$? To answer this question, we first suppose that $r_{ij} = 1$ for all $i$ and $j$. For $a_{ii} = w_{ii} = 0$ and $r_{ij} = 1$, we have 
\begin{equation}
    \bar{s}_i = \frac{1}{k_i}\sum_{j=1}^N a_{ij} e^{\ii \varphi_{ij}}\,,
\label{eq:s_i}
\end{equation}
so
\begin{align}
\begin{split}
    |\bar{s}_i|^2&=\frac{1}{k_i^2}\sum_{j,l}  a_{ij} a_{il} e^{\ii (\varphi_{ij}-\varphi_{il})}\\
    &=\frac{1}{k_i^2}\sum_{j,l} a_{ij} a_{il}\cos(\varphi_{ij}-\varphi_{il})\,.
\end{split}
\label{eq:s_i2}
\end{align}
Observe that $|\bar{s}_i|^2 = 1$ if $\varphi_{ij} = \varphi_{il}$ for all edges that are attached to node $i$ and that $\bar{s}_i$ reaches a minimum if all phases are balanced around the unit circle (\eg, if they are spread evenly or distributed in clusters that balance each other). For phase distributions other than these two cases, $|\bar{s}_i|^2$ satisfies $0 < |\bar{s}_i|^2 < 1$. When $r_{ij} = 1$, Eq.~\eqref{eq:s_i2} implies that the squared magnitude of the normalized node strength is similar to the order parameter of the Kuramoto model [see Eq.~\eqref{eq:r_q}].

If $r_{ij} = r$, then $0 < |\bar{s}_i|^2 < r^2$. Additionally, for general distributions of $r_{ij}$ with $r_{\rm max}=\max_{i,j} (r_{ij})$, {the quantity $|\bar{s}_i|^2$ satisfies} $0<|\bar{s}_i|^2<r_{\rm max}^2$.

Generalizing the definition of weighted nearest-neighbor degree from Ref.~\cite{barrat2004architecture} to networks with complex edge weights yields
\begin{equation}
    k_{{\rm nn},i}^w = \frac{1}{s_i}\sum_{j=1}^N w_{ij} k_j\,.
\end{equation}
One can separately track real and imaginary nearest-neighbor degrees by calculating $\Re(k_{{\rm nn},i}^w)$ and $\Im(k_{{\rm nn},i}^w)$.
\begin{figure}
    \centering
    \includegraphics[width=0.48\textwidth]{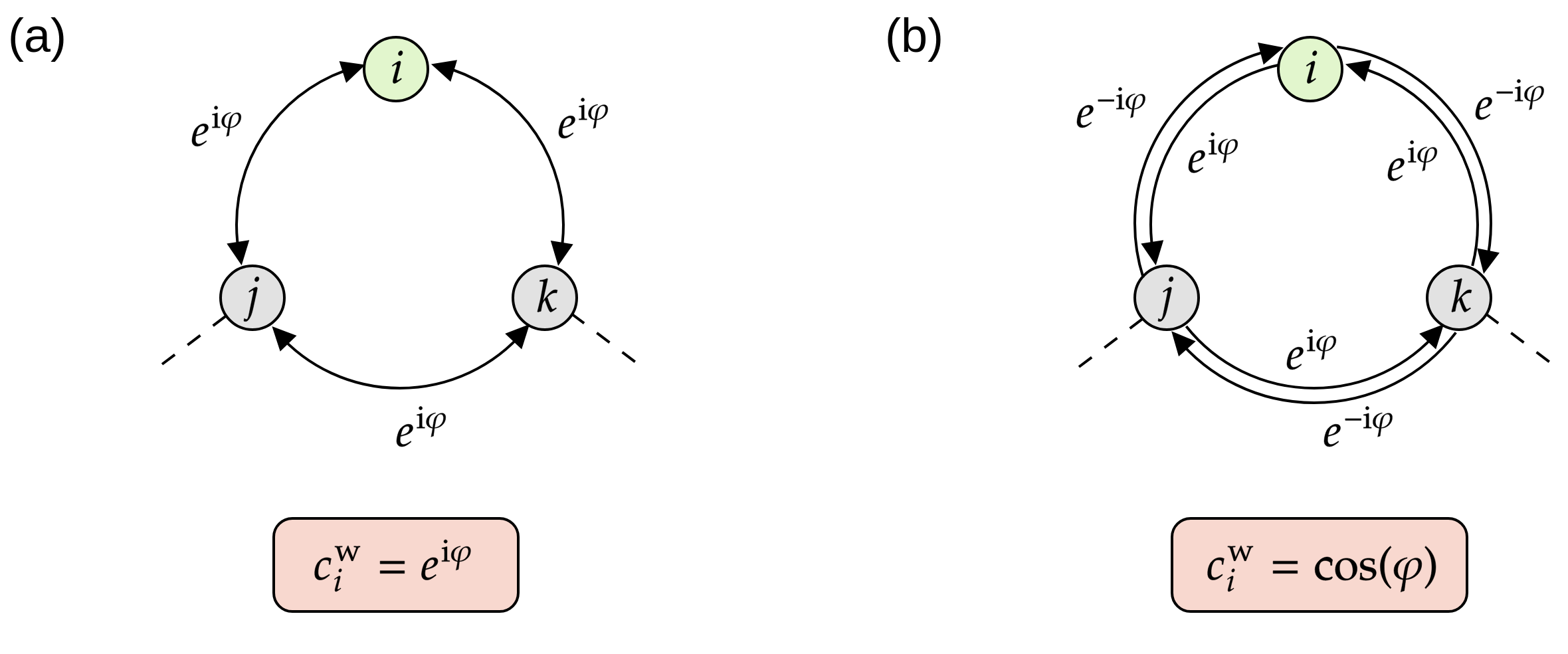}
    \caption{\textbf{Examples of clustering coefficients in a network with complex edge weights.} (a) A closed and directed triad with complex weights $e^{\ii \varphi}$. The weighted clustering coefficient of node $i$ (in green) is $c_i^{\rm w}=e^{\ii \varphi}$. (b) A closed and directed triad with complex weights $e^{\pm \ii \varphi}$. The weighted clustering coefficient of node $i$ (in green) is $c_i^{\rm w}=\cos(\varphi)$.}
    \label{fig:clustering}
\end{figure}

In a binary and undirected network, the local clustering coefficient~\cite{cozzo2015structure} of node $i$ is
\begin{equation}
    c_i=\frac{1}{k_i(k_i-1)}\sum_{j,k}a_{ij}a_{jk}a_{ki}
\label{eq:ci}
\end{equation}
when $k_i \geq 2$. For $k_i = 0$ and $k_i = 1$, we set $c_i=0$.
There are a variety of ways to define local clustering coefficients in weighted networks~\cite{barrat2004architecture,onnela2005intensity,saramaki2007generalizations}. Drawing inspiration from Ref.~\cite{onnela2005intensity}, we define the local weighted clustering coefficient of an undirected network with complex weights as
\begin{equation}
    c_i^{\rm w}=\frac{1}{k_i(k_i-1)}\sum_{j,k}(\tilde{w}_{ij} \tilde{w}_{jk} \tilde{w}_{ki})^{1/3}\,,
\label{eq:ci_weighted}
\end{equation}
where $\tilde{w}_{ij}=w_{ij}/\max_{i,j}{|w_{ij}|}$ is the normalized weight of the edge 
between nodes $i$ and $j$. We also use Eq.~\eqref{eq:ci_weighted} for directed networks in which the in-degree of each node is equal to its out-degree. (For more general directed networks, it is necessary to further generalize Eq. \eqref{eq:ci_weighted}.) For $w_{ij}=r a_{ij}$ with $r>0$, this weighted clustering coefficient equals the unweighted clustering coefficient $c_i$. 

Instead of counting all triangles (\ie, ``closed'' triads, in which all possible edges are present) that are associated with a certain node in the same way, the weighted clustering coefficient $c_i^{\rm w}$ accounts for differences in the edge weights. For example, if a triangle connects nodes $i$, $j$, and $k$, then the unweighted local clustering coefficient counts the corresponding edges while ignoring their weights.
However, if all of the normalized weights $\tilde{w}_{ij}$ that are associated with that triangle are close to $0$, one may wish to 
weight the triangle differently than other (more important) triangles with larger edge weights. In networks with complex weights, it is possible to account not only for positive and negative edges (which arise, \eg, in correlation networks~\cite{costantini2014generalization,masuda2018clustering} and in subjects such as international relations~\cite{cartwright1956structural,marvel2011continuous}), but also to quantify directional and phase information (in addition to magnitudes). In Fig.~\ref{fig:clustering}(a), we show an example of a closed and undirected triad with edge weights $e^{\ii \varphi}$. The local weighted clustering coefficient of node $i$ is $c_i^{\rm w}=e^{\ii \varphi}$. In the triad in Fig.~\ref{fig:clustering}(b), we use the edge weights $e^{\pm \ii \varphi}$ to encode directional information. The corresponding local weighted clustering coefficient of node $i$ is $c_i^{\rm w} = \cos(\varphi)$. This example illustrates that one can use the weighted clustering coefficient $c_i^{\rm w}$ to characterize the local distribution of complex edge weights. If all weights have the same magnitude but the phases that are associated with the two cycles $i\rightarrow j\rightarrow k \rightarrow i$ and $i\rightarrow k\rightarrow j \rightarrow i$ have opposite signs, then the imaginary part of $c_i^{\rm w}$ is $0$.
%

%%%%%%

\section{Matrix powers and walks}
\label{sec:matrix_powers_walks}
Given an adjacency matrix $A$, the entries $a_{ij}^{(k)}$ of the matrix powers $A^k$ (with $k \in \{1,2,\ldots\}$) correspond to the number of walks of length $k$ that start at node $i$ and end at node $j$~\cite{estrada2012structure}. For a weight matrix $W$, each entry $w_{ij}^{(k)}$ of the matrix power $W^{k}$ is equal to the sum of the products of all weights that are associated with length-$k$ walks from node $i$ to node $j$. If $W$ is a stochastic matrix, the entries of $W^{k}$ correspond to the probabilities of reaching certain nodes from other nodes. 

One advantage of complex-valued weights over real-valued weights is that 
using complex values allows one to encode directional information. Consider the network in Fig.~\ref{fig:complex_weights}(b). 
Clockwise and counterclockwise walks on this network have negative and positives phases, respectively. One can determine the ``direction'' of a walk in such a network from the accumulated phase of the product of the weights that are associated with edges that a walker traverses. For example, the walks $1\rightarrow 2 \rightarrow 1$ and $1\rightarrow 2 \rightarrow 3$ have total weights of $e^{\ii 0}$ and $e^{\ii 2 \pi/3}$, respectively. Therefore, their phases $0$ and $2\pi/3$ indicate that the first walker traverses edges of opposite phase and returns to its initial position and that the second walker moves counterclockwise.
In a quantum picture, one can interpret the positive and negative phases that are associated with a walk $i\rightarrow j$ as Aharonov--Bohm phases that result from interactions between a charged particle and a magnetic vector potential $\mathbf{A}$~\cite{avron1988adiabatic,Lieb2004}. That is,
\begin{equation}
    \varphi_{ij}=\int_{\mathcal{C}_{ij}}\mathbf{A}\cdot \mathrm{d}\mathbf{x}\,,
\end{equation}
where $\varphi_{ij}=-\varphi_{ji}$ and $\mathcal{C}_{ij}$ is a curve that starts at position $\mathbf{x}_i$ and ends at position $\mathbf{x}_j$. The amplitude $r_{ij}$ in $w_{ij}=r_{ij} e^{\ii \varphi_{ij}}$ is equal to $a_{ij}$. That is, $r_{ij} = 1$ if the charged particle can move from $\mathbf{x}_i$ to $\mathbf{x}_j$ and otherwise  $r_{ij} = 0$. The resulting weight matrix is a discrete version of the magnetic Laplacian~\cite{Lieb2004}.

The complex edge weights  $w_{ij}=e^{\ii\pi/3}=(1+\ii \sqrt{3})/2$ and $w_{ji}=\bar{w}_{ij}=e^{-\ii\pi/3}=(1-\ii \sqrt{3})/2$ in the triads in Fig.~\ref{fig:complex_weights}(b) have useful properties for applications in network analysis because both their product and sum are equal to $1$ (\ie, $w_{ij} w_{ji}=1$ and $w_{ij}+w_{ji}=1$)~\cite{mohar2020new}. In the associated weight-matrix powers, a walk $i\rightarrow j \rightarrow i$ contributes $1$ just as in the corresponding adjacency-matrix powers. As an example, we again consider the network in Fig.~\ref{fig:complex_weights}(b) and calculate the square of its weight matrix $W$ and adjacency matrix $A$. We obtain
\begin{equation}
    W^2 = \begin{pmatrix}
2 & e^{-\ii 2\pi/3} & e^{\ii 2\pi/3} \\
e^{\ii 2\pi/3} & 2 & e^{-\ii 2\pi/3} \\
e^{-\ii 2\pi/3} & e^{\ii 2\pi/3} & 2
\end{pmatrix}\,,\,
A^2 = \begin{pmatrix}
2 & 1 & 1 \\
1 & 2 & 1 \\
1 & 1 & 2
\end{pmatrix}\,.
\label{eq:matrix_power}
\end{equation}
In this example, the diagonal entries of $W^2$ and $A^2$ are equivalent 
because of the multiplicative and additive properties of $e^{\pm \ii\pi/3}$.

Complex weights with fractional phases such as $\varphi = \pi/3$ arise in quantum-mechanical particle statistics. 
%While e
Exchanging bosonic particles in a multi-particle wave function is associated with a phase $\varphi = 0$, and exchanging fermions is associated with the phase
%the exchange phase of fermions is 
$\varphi=\pi$. One can realize fractional phases $\varphi=\pi/(2m+1)$ (with $m\in\mathbb{N}_{>0} = \{1, 2, \ldots \}$) that lie between those of bosons and fermions using anyons, which are quasiparticles that arise in 2D systems~\cite{leinaas1977theory,wilczek1982magnetic,wilczek1982quantum,abelian_anyons2020,nakamura2020direct}. 
The phases $\varphi=\pi/3$ have been observed experimentally in anyon systems~\cite{abelian_anyons2020,nakamura2020direct}. In a network that represents anyon permutations, one obtains the total phase that is associated with a permutation by multiplying the complex weights $e^{\ii \varphi_{ij}}$ in the $U(1)$ representation of the underlying braid group~\cite{macikazek2019non}. In an anyon system with exchange phase $\varphi=\pi/3$, exchanging two particles twice is associated with a total phase of $\varphi=2\pi/3$, as one can see in some of the off-diagonal entries of $W^2$ in Eq.~\eqref{eq:matrix_power}.

%%%%%%%%%

%
\section{Graph energy}
\label{sec:graph_energy}
\begin{figure}
    \centering
    \includegraphics{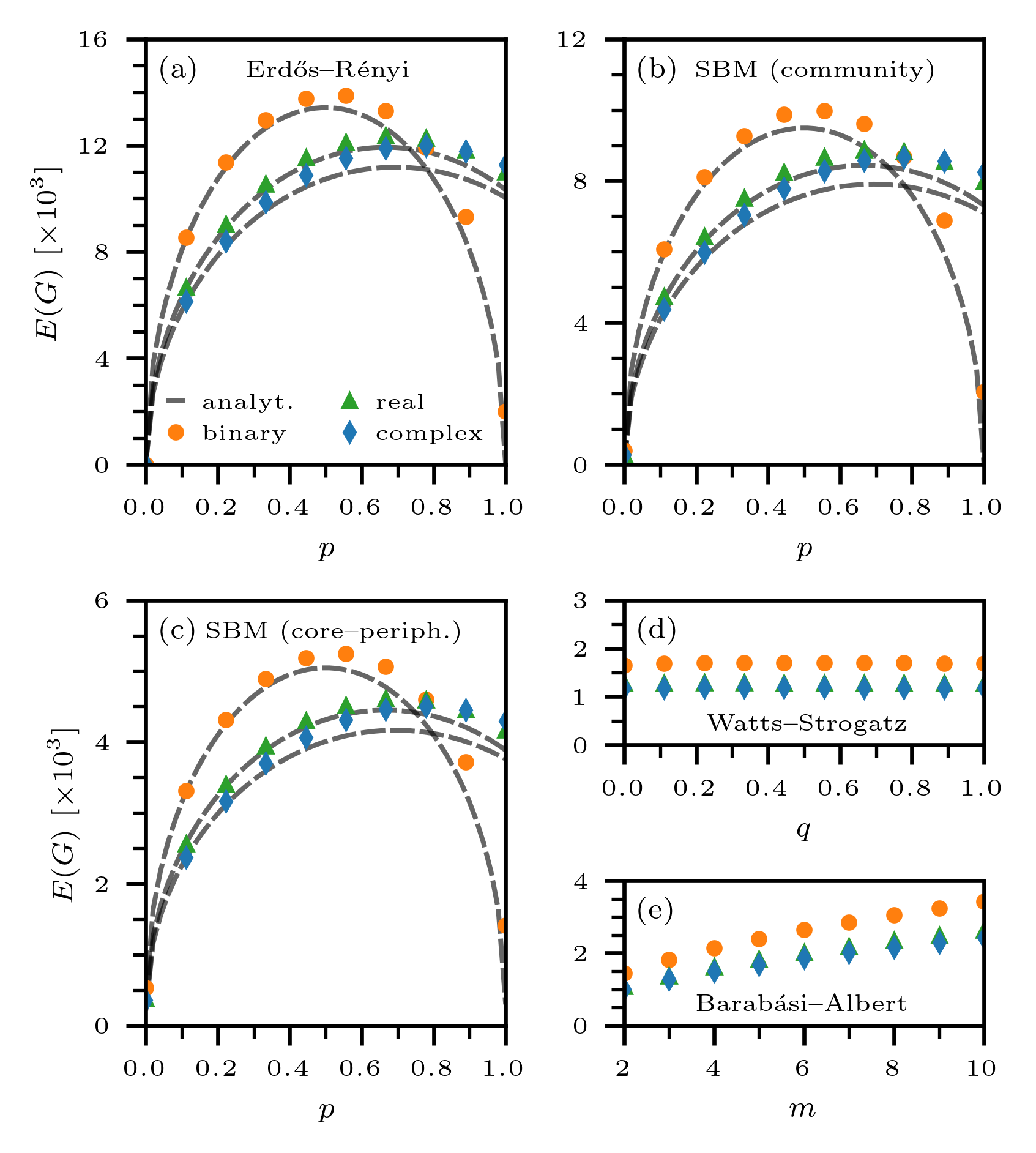}
    \caption{\textbf{Graph energies for different networks with binary, real, and complex edge weights.}
     We show the graph energies $E(G)$ [see Eq.~\eqref{eq:energy}] for five types of networks with binary, real, and complex weight distributions:
         (a) a $G(N,p)$ ER network, (b) a stochastic block model (SBM) with two $G(N,p)$ ER blocks and inter-block connection probability $10^{-3}$, (c) an SBM with one $G(N,p)$ ER block, one $G(N,p)$ ER block with $p = 10^{-3}$, and inter-block connection probability $10^{-3}$, (d) a $G(N,k,q)$ Watts--Strogatz (WS) network in which each node is adjacent to $k = 4$ nearest neighbors (where $q$ denotes the probability of rewiring each edge), and (e) a Barab\'asi--Albert (BA) network.      
    All of these networks have $N = 1000$ nodes. In all simulations that involve weighted networks, we use Hermitian weight matrices (\ie, $W=W^\dagger$). To construct the BA network, we start with a star graph with $1$ hub and $m$ leaves, and we iteratively add new nodes until there are $1000$ nodes. Each new node has $m$ edges that connect to existing nodes using linear preferential attachment. The orange disks indicate numerical results for binary weight matrices (\ie, for $W = A$). The green triangles indicate numerical results for networks with real-valued weight distributions, and the blue diamonds indicate numerical results for networks with complex-valued weight distributions.
     The real weights are distributed uniformly in the interval $[0,4/3)$, and the complex weights are distributed uniformly in the first quadrant of the complex plane. 
     All reported results are means of $100$ independent instantiations of the indicated random-graph models.  
For each instantiation, we use the same network structure, but we change the weights (which can be binary, real, or complex).
    The dashed gray curves in panels (a)--(c) are based on the analytical solutions \eqref{eq:ER_energy}, \eqref{eq:graph_energy_weighted}, and \eqref{eq:graph_energy_weighted_complex}, which assume that $N \rightarrow \infty$.}
    \label{fig:graph_energy}
\end{figure}
%

%%%%

The energy of a graph is 
\begin{equation}
    E(G)=\sum_{i=1}^N |\lambda_i|\,,
\label{eq:energy}
\end{equation}
where $\lambda_i$ is the $i$th eigenvalue of the weight matrix $W$~\cite{li2012graph}. 
To gain insight into the differences in the energy of
binary networks, networks with real edge weights, and networks with complex edge weights, we compute 
graph energy for several well-studied types of networks.

In H\"uckel molecular-orbital (HMO) theory (\ie, tight-binding molecular-orbital theory), one typically represents conjugated hydrocarbon molecules by undirected and binary networks. The energy of $\pi$-electrons in this HMO approximation is equivalent to the energy in Eq.~\eqref{eq:energy}~\cite{li2012graph}. Although most applications of network analysis in mathematical chemistry have focused on undirected and binary molecular networks, weight matrices with real and complex entries have also been studied~\cite{estrada2006atomic} (\eg, to examine
cis/trans isomers of molecules~\cite{lekishvili1997characterization,golbraikh2002novel}). 

One can obtain closed-form expressions for the graph energy of certain graphs~\cite{li2012graph}. For example, the energy of almost every\footnote{In this paper, we say that ``almost every'' graph $\hat{G}$ in a random-graph model $G$ with $N$ nodes has a certain property if the probability that $\hat{G}$ satisfies that property approaches $1$ as $N \rightarrow \infty$.} $G(N,p)$ ER network~\cite{erdos59a,li2012graph} is
\begin{equation}
    E[G(N,p)]=N^{3/2}\left(\frac{8}{3\pi}\sqrt{p(1-p)}+\smallO(1)\right)\,,
\label{eq:ER_energy}
\end{equation}
where $p$ is the connection probability. There are very few energy estimates for weighted networks, so in particular there are few such estimates for networks with complex edge weights~\cite{liu2015hermitian}. We use graph energy to characterize ER, stochastic-block-model (SBM), Watts--Strogatz (WS)~\cite{watts1998collective}, and Barab\'asi--Albert (BA)~\cite{barabasi1999emergence} networks with binary, real, and complex weight distributions (see Fig.~\ref{fig:graph_energy}). We study one 2-block SBM with community structure and one 2-block SBM with core--periphery structure. 
In our calculations, we distribute the complex weights uniformly at random in the subset of the unit circle in the first quadrant of the complex plane [\ie, $\varphi_{ij} \sim \mathcal{U}[0,\pi/2)$ and $r_{ij} = \sqrt{\epsilon_{ij}}$, where $\epsilon_{ij} \sim \mathcal{U}[0,1)$ and $\mathcal{U}[a,b)$ denotes the uniform distribution on the half-open interval $[a,b)$] and distribute the real weights uniformly at random in the interval $[0,4/3)$ [\ie, $w_{ij} \sim \mathcal{U}([0,4/3))$]. 
We use the interval $[0,4/3)$ to ensure that the mean value of the real weights is equal to the mean of the absolute value of the complex weights. The examined weight matrices are Hermitian. In App.~\ref{app:eigenvalue_distr}, we derive analogues of Eq.~\eqref{eq:ER_energy} for ER networks with these two weight distributions.

In Fig.~\ref{fig:graph_energy}, we show sample means of the graph energy $E(G)$ for ER, SBM, WS, and BA networks as a function of the parameters of these random-graph models.
We show the corresponding eigenvalue distributions in App.~\ref{app:eigenvalue_distr}. For an ER network with binary edges [see Fig.~\ref{fig:graph_energy}(a)], the maximum of graph energy as $N \rightarrow \infty$ occurs when $p = 0.5$ [see Eq.~\eqref{eq:ER_energy}], whereas we observe that the examined real and complex weight distributions have graph-energy maxima when $p\approx 0.8$. 
The largest difference in graph energy between the examined ER networks with binary edges and their counterparts with real and complex edge weights occur for {$p = 1$} (\ie, in a fully connected graph). A binary ER network has a graph energy of $E[G(N,p=1)] = 2(N-1)$ (see App.~\ref{app:ER_graph_energy}), whereas the corresponding weighted networks have significantly larger values of graph energy for $p=1$ and $N = 1000$. The graph energies of the examined SBM networks, in which we vary only a single probability parameter, have similar qualitative behavior as the ER networks [see Fig.~\ref{fig:graph_energy}(b,c)]. 
For small inter-block connection probabilities, one can approximate the graph energy of an SBM network that consists of ER blocks by the sum of the corresponding ER graph energies \eqref{eq:ER_energy}.

The mean graph energy of the WS networks is largely independent of the rewiring probability $q$. In the BA networks, the mean graph energy increases with $m$, which is the number of new edges that one adds for each new node. For the examined WS and BA networks, the mean graph energies that we obtain for binary weights are larger than those for the examined real and complex weights. Although the mean values of the examined real weights equal the means of the absolute values of the associated complex weights, the mean values of $E(G)$ for the ER networks with real weights are larger than those of their counterparts with complex weights for $p\lessapprox 0.8$. For $p \gtrapprox 0.8$, the mean graph energy is larger for ER networks with complex weights than for ER networks with real or binary weights. In the WS and BA networks, the graph energies that we obtain with real weights are about 5\% and 8\% larger, respectively, than those with complex weights.

In summary, graph energy is qualitatively different in binary
networks, networks with real edge weights, and networks with complex edge weights. This is the case both for the magnitude of the graph energy and for how it depends on the network connectivity properties 
(as quantified by the network parameters $p$, $q$, and $m$). Given the described connections between graph energy and energy estimates of conjugated molecules, our results may be relevant to the modeling of molecules with weighted networks.
%
%%%%%%%%

\section{The Perron--Frobenius theorem and eigenvector centrality}
\label{sec:pf_eig_cen}

According to the Perron--Frobenius theorem~\cite{perron1907,frobenius1912}, the weight matrix $W = A$ of a binary, strongly connected network is associated with a simple positive eigenvalue (the ``Perron eigenvalue'') that is strictly larger than all other eigenvalues. The Perron eigenvalue is equal to the spectral radius of $W$. That is, $\rho(A)\coloneqq \max_{i \in \{1, \ldots,N\}}
\{|\lambda_i|\}$, where $\{\lambda_i\}_{i \in \{1, \ldots, N\}}$ is the set of eigenvalues (\ie, the spectrum) of $W$. The corresponding Perron eigenvector (\ie, the leading eigenvector) is a centrality measure~\cite{newman2018networks}.

The Perron--Frobenius theorem does not hold for matrices with complex weights, so we cannot find a Perron eigenvalue of $W$ and a corresponding eigenvector to use as a centrality measure. However, generalizations of the Perron--Frobenius theorem~\cite{rump2003perron,noutsos2012perron} provide a possible approach to define eigenvector centrality (and generalizations of it, such as PageRank) for networks with certain types of complex weight matrices. A complex weight matrix $W$ has the ``strong Perron--Frobenius property'' if it (1) has a simple positive eigenvalue $\lambda_1$ that satisfies $\lambda_1 = \rho(W) > |\lambda_i|$ (with $i \in \{2, \ldots, N\}$) and (2) has a corresponding column eigenvector $v_1$ with positive entries. The eigenvector $v_1$ is called the ``right Perron--Frobenius eigenvector''. For further details about generalizations of the Perron--Frobenius theorem to complex matrices, see Ref.~\cite{noutsos2012perron}.

An example of a weight matrix with the strong Perron--Frobenius property is
\begin{equation} \label{eq:strong_PF}
	W=
		\begin{pmatrix}
			0 & a e^{\ii \varphi_a} & 0 \\
			b e^{\ii \varphi_b} & 0 & c e^{\ii \varphi_c} \\
			0 & d e^{\ii \varphi_d} & 0
		\end{pmatrix}
\end{equation}
with $a=b=c=1$, $d=-1/2$, $\varphi_a=0$, $\varphi_b=\pi/6$, $\varphi_c=3\pi/2$, and $\varphi_d=\pi$. Inserting these parameter values gives
\begin{equation} \label{eq:strong_PF_2}
	W=
		\begin{pmatrix}
			0 & 1 & 0 \\
			e^{\ii \pi/6} & 0 & -\ii \\
			0 & 1/2 & 0
		\end{pmatrix} \,.
\end{equation}
The largest eigenvalue of $W$ is $\sqrt[4]{3/4}$, and its corresponding eigenvector is $v_1=(2,\sqrt[4]{12}, 1)^\top$. What is the meaning of the right Perron--Frobenius vector $v_1=(2,\sqrt[4]{12}, 1)^\top$ in the context of the eigenvector centrality of a complex weight matrix? In this example, nodes $1$, $2$, and $3$ have eigenvector centralities of $2$, $\sqrt[4]{12}$, and $1$, respectively. The most central nodes in the associated network are thus nodes $1$ and $2$, whereas node $3$ (which has a single out-edge with weight $1/2$) is the least central node.
In this example, the eigenvector centralities that we obtain using the complex weight matrix $W$ coincide with the eigenvector centralities that we obtain from $\Re(W)$. 

In App.~\ref{app:strong_perron_frobenius}, we show that any Hermitian two-node network with complex edges weights with nonzero imaginary parts cannot satisfy the strong Perron--Frobenius property.
%
%%%%%%

\section{Betweenness and closeness centralities}
\label{sec:bet_clo_cent}
\begin{figure}
  \centering
  \includegraphics[valign=m,width=0.2\textwidth]{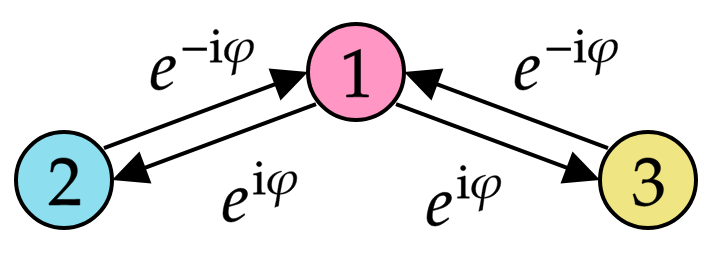}
  \qquad
  \renewcommand*{\arraystretch}{1.5}
  \begin{tabular}{lccc}\toprule
\multicolumn{1}{c}{} & \multicolumn{1}{c}{\textbf{1}} & \multicolumn{1}{c}{\,\,\textbf{2}} & \multicolumn{1}{c}{\textbf{3}} \\\hline
\,\,\, betweenness  \,\,\,& 1 & \,\,0 & \,0 \\
\,\,\, closeness  \,\,\,& 1 & \,\,2/3 & \,2/3 \\\bottomrule
\end{tabular}
 \caption{{\textbf{Betweenness and closeness centralities of a small
 network.}} A network with three nodes and the weight matrix \eqref{eq:toy_example}. In the table, we show the geodesic betweenness and closeness centralities of nodes $1$, $2$, and $3$ that are associated with the unweighted (\ie, binary)
 analogue of the depicted network.}
\label{fig:toy_example}
\end{figure}
\begin{figure}
    \centering
    \includegraphics{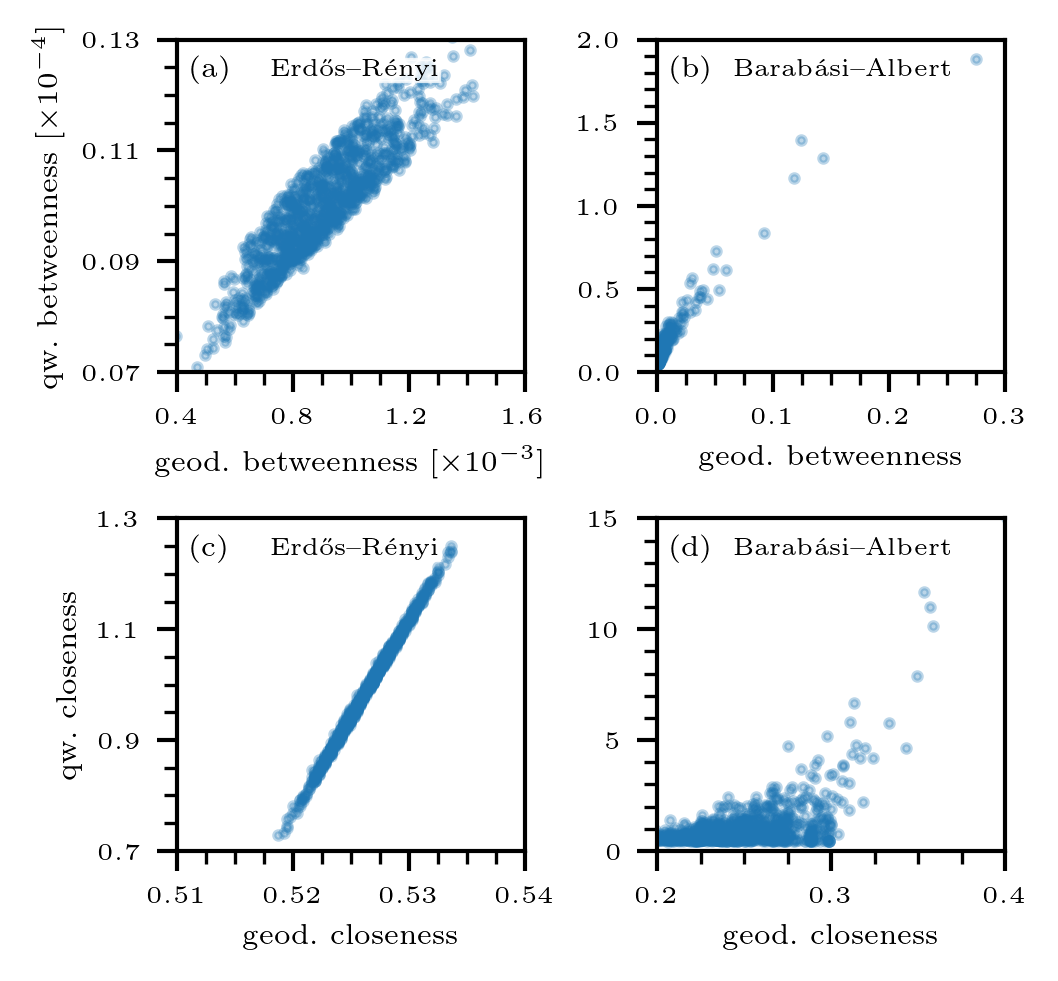}
    \caption{\textbf{Quantum random-walk versus geodesic betweenness and closeness centralities.} We illustrate the correlations between quantum random-walk (qw) and geodesic centralities for (a,b) betweenness and (c,d) closeness. In (a,c), we show scatter plots for a $G(N,p)$ ER network with $p = 0.1$. In (b,d), we show scatter plots for a BA network that we construct from an initial star graph with $1$ hub
    and $2$ leaves by iteratively adding new nodes until there are $N = 1000$ nodes. 
    Each new node has $m = 2$ edges that connect to existing nodes using linear preferential attachment. Both networks have $1000$ nodes. The Pearson correlation coefficients are (a) 0.89, (b) 0.81, (c) 1.00, and (d) 0.36. We calculate the geodesic centralities for binary
    networks. To calculate the quantum random-walk centralities, we use $H_{\rm q} = -W$ [see Eq.~\eqref{eq:CTQW}] as the evolution operator. We suppose that $W$ is Hermitian, so
    $W = W^\dagger$. Aside from the Hermiticity constraints, we distribute the weights uniformly at random in the first quadrant of the complex plane.
    } 
    \label{fig:centralities}
\end{figure}
The geodesic betweenness centrality $c_{\rm B}(i)$ of a node $i$ quantifies the number of shortest paths that traverse that node, and the closeness centrality $c(i)$ of a node $i$ quantifies the mean distance between that node and other nodes~\cite{newman2018networks}. Mathematically, the normalized geodesic betweenness centrality of node $i$ in a directed network is
\begin{equation}
    c_{\rm B}(i) = \frac{1}{(N-1)(N-2)} \mathlarger{\mathlarger{\sum}}_{\{j,k | j\neq i, k \neq i\}}\frac{\sigma_{jk}(i)}{\sigma_{jk}}\,,
\end{equation}
where $\sigma_{jk}$ is the total number of shortest paths between nodes $j$ and $k$ and $\sigma_{jk}(i)$ is the number of those shortest paths that traverse node $i$. The closeness centrality of node $i$ is
\begin{equation}
    c(i) = \frac{N-1}{\sum_{j\neq i} d_{ij}}\,,
\end{equation}
where $d_{ij}$ is the geodesic (\ie, shortest-path) distance between nodes $i$ and $j$.

Because complex numbers are not fully ordered, one cannot use geodesic betweenness and closeness centrality measures that are based on shortest paths on networks with complex weights. Except in degenerate situations, one cannot order path lengths in networks with complex edge weights. As in the above examples for occupation and eigenvector centralities (see Secs.~\ref{sec:compl_weights} and \ref{sec:pf_eig_cen}), one has to appropriately generalize geodesic betweenness and closeness centralities. One way to quantify node importance in a network with a complex weight matrix is to use quantum random-walk centrality measures~\cite{boettcher2021classical}. In Eq.~\eqref{eq:quant_occ}, we give an occupation centrality measure that is based on a quantum random walk with Hamiltonian $H_{\rm q} = -W$ and Hermitian $W$. In App.~\ref{app:betweenness_closeness}, we describe corresponding generalizations of betweenness and closeness that are based on absorbing random walks and can take complex weight matrices as inputs. 

We first apply these centrality measures to a network with three nodes (see Fig.~\ref{fig:toy_example}) and weight matrix
\begin{equation} \label{eq:toy_example}
   	 W = \begin{pmatrix}
			0 & e^{\ii \varphi} & e^{\ii \varphi} \\
			e^{-\ii \varphi} & 0 & 0 \\
			e^{-\ii \varphi} & 0 & 0
		\end{pmatrix}\,.
\end{equation}
In the associated unweighted (\ie, binary) analogue of this network, the geodesic betweenness centralities of nodes 1, 2, and 3 are 1, 0, and 0, respectively. The corresponding quantum random-walk betweenness centralities that are associated with the Hamiltonian $H_{\rm q} = -W$ with $\varphi=\pi/3$ are $1$, $0.65$, and $0.65$. (We normalize the betweenness values so that the maximum is $1$.) The ranking of the nodes is the same as with geodesic betweenness centrality. To calculate a quantum random-walk version of closeness centrality (see App.~\ref{app:betweenness_closeness}), we first calculate the mean return time (\ie, the inverse of the occupation probability). We calculate quantum random-walk occupation centrality $\pi_j$ (with $j \in \{1,2,3\}$) by evaluating the infinite-time mean \eqref{eq:quant_occ}, where $\ket{\psi(t)}=e^{-\ii H_{\rm q} t}\ket{\psi(0)}$. We set $\ket{\psi(0)} = (1,1,1)^\top/\sqrt{3}$ and find that the infinite-time mean is $\boldsymbol{\pi}=(0.5,0.25,0.25)^\top$. In this example, the occupation centrality of node 1 is twice as large as that of nodes 2 and 3. The corresponding quantum random-walk closeness values are $1$, $0.9$, and $0.9$. 
(We normalize these values so that the maximum is $1$.) As with betweenness centrality, we find that quantum random-walk closeness yields the same node ranking as geodesic closeness.

We now compare geodesic and quantum random-walk centralities for ER and BA networks with $N = 1000$ nodes and complex edge weights that (as in our examination of graph energy) we distribute uniformly at random in the subset of the unit circle in the first quadrant of the complex plane. 
We again consider the evolution operator $H_{\rm q} = -W$ [see Eq.~\eqref{eq:CTQW}]. In Fig.~\ref{fig:centralities}, we show scatter plots to compare the geodesic and quantum random-walk 
centralities. 
In the examined ER and BA networks, both betweenness and closeness have Pearson correlation coefficients that range from $0.36$ to $1.00$.
Our results suggest that quantum random-walk closeness and betweenness centralities are able to rank node importance in networks with complex weights in a manner that is similar to their corresponding geodesic centralities.
%

%%%%%%

\section{Conclusions and discussion}
\label{sec:disc_concl}
Networks with complex-valued edge weights arise in a variety of situations. However, most studies of weighted networks have focused primarily on networks with real-valued edge weights. 

In the present paper, we examined network-analysis methods that are useful to study the structure of networks with complex edge weights. To physically interpret such networks and the underlying directional information that is encoded in the phases of the complex weights, we discussed connections between complex weight matrices and salient physical systems. For example, perhaps the phases that are associated with walks in a network with complex edge weights arise from 
interactions between a charged particle (which traverses the edges) and a vector potential.
Moreover, akin to the interpretation of stochastic weight matrices as generators of linear diffusion dynamics (\ie, random walks), we showed that one can interpret Hermitian weight matrices with complex entries as generators of time translation in continuous-time quantum walks. 
We also generalized the DeGroot model of consensus dynamics to networks with complex edge weights. Finally, we characterized the structural features of networks with complex edge weights using a variety of network measures (specifically, graph energy, common centrality measures, and generalizations of node strength and a local clustering coefficient).

Given the diverse variety of applications of networks with complex weights (see Tab.~\ref{tab:examples}), there are many interesting directions for future work. We mention a few of them in passing. A very recent paper examined random walks and structural balance in networks with complex edge weights \cite{tian2023}, and there are many exciting directions to pursue to build on it. Another potentially valuable area is investigating the properties of electrical networks of resistors, coils, and capacitors~\cite{alonso2017power,chen2017power,muranova2019notion,muranova2020eigenvalues,muranova2021effective,muranova2022effective,muranova2022networks} using network-analysis approaches. In such networks, one can use complex weights to describe complex impedances and reflection coefficients. Another avenue for future research is an analytical investigation of graph energy in networks with both real and complex weights. Such efforts can build fruitfully on random-matrix-theory analyses of weighted networks~\cite{baron2022eigenvalue}. Furthermore, given the importance of the Perron--Frobenius theorem for dynamical processes and centrality measures in networks with real weights, it is worthwhile to identify physical systems with associated complex weight matrices that satisfy the strong Perron-Frobenius property. 
Studies of such systems may help further guide the development of suitable centralities, spectral clustering methods~\cite{michoel2012alignment}, and other network measures to study physical systems with complex weight matrices.
Additionally, given the relevance of motifs in the study of both unweighted and weighted networks, it seems worthwhile to examine motifs in networks with complex edge weights, such as by generalizing walk-based motifs from classical contexts~\cite{schwarze2021} to quantum ones.
%
%%%%%

\acknowledgements
We thank Ginestra Bianconi, Tom Burns, Lincoln Carr, Karen Daniels, Alexander Goltsev, Malte Henkel, Adam Knapp, Cris Moore, Matteo Paris, and two anonymous referees for helpful comments.
%
%%%%%%%%%
%
\section*{Data and code availability}
The code and data that support the findings of the present study are publicly available at \url{https://gitlab.com/ComputationalScience/complex-weights}.
\appendix
\section{Quantum synchronization}
\label{app:quantum_synch}
The quantum-state evolution \eqref{eq:quantum_consensus} conserves the $L_2$ norm that is associated with $\ket{\psi_i}$. That is, $\partial_t \|\psi_i\|^2=\braket{\partial_t \psi_i|\psi_i}+\braket{\psi_i|\partial_t \psi_i}=0$ because
\begin{align}
    \braket{\psi_i|\partial_t \psi_i} &= \sum_{j=1}^N a_{ij} \left[\braket{\psi_i|\psi_j}-\braket{\psi_j|\psi_i}\right] \,, \notag \\
    \braket{\partial_t \psi_i|\psi_i} &= \sum_{j=1}^N a_{ij} \left[\braket{\psi_j|\psi_i}-\braket{\psi_i|\psi_j}\right]\,.
\end{align}
%

%%%%%%

\section{Eigenvalue distributions}
\label{app:eigenvalue_distr}
\begin{figure}
    \centering
    \includegraphics{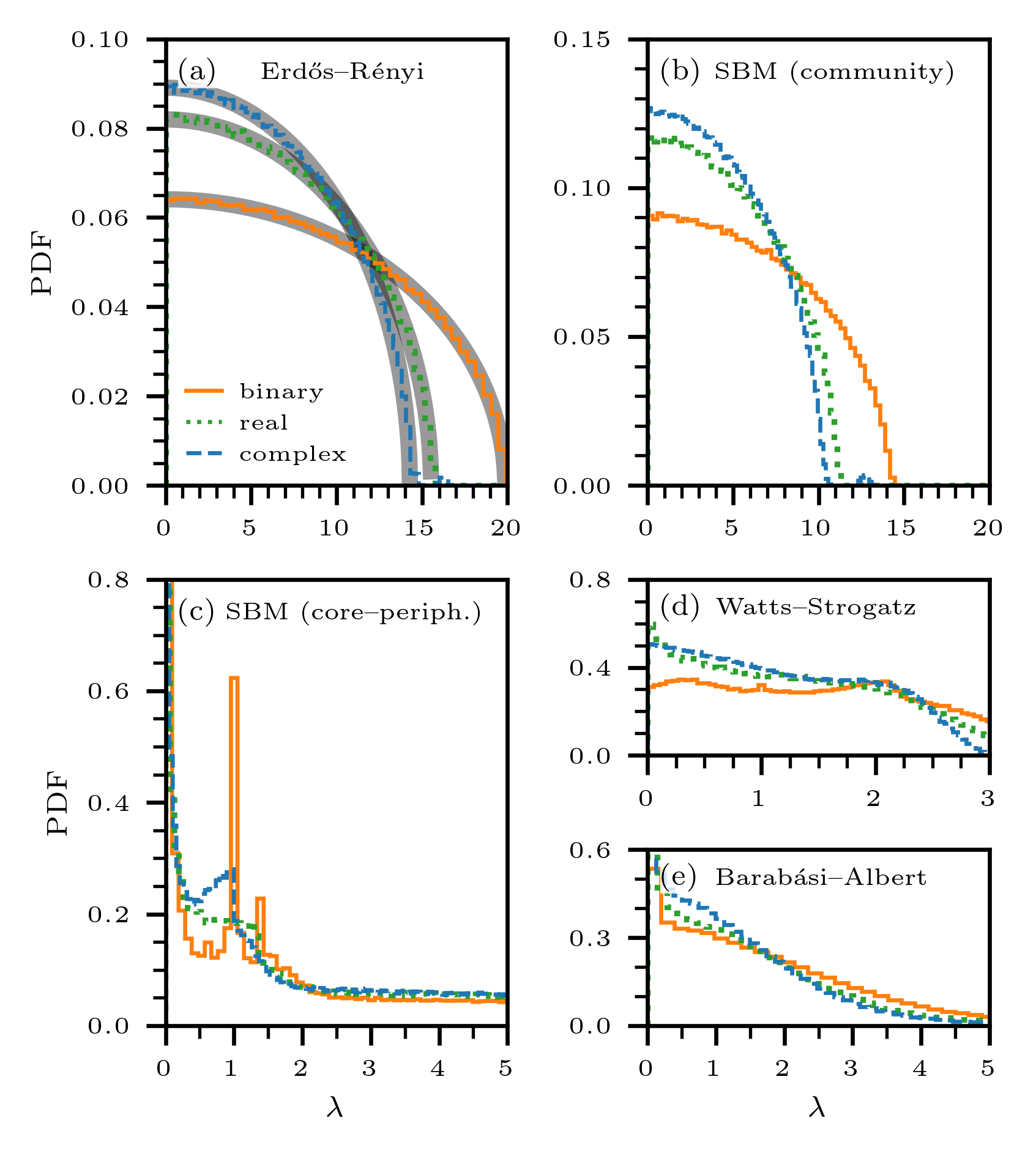}
    \caption{\textbf{Eigenvalue distributions for different networks with binary, real, and complex edge weights.}
     %and weight distributions.} 
     We show the eigenvalue distributions for five types of networks with binary, real, and complex weight distributions: (a) a $G(N,p)$ ER network with $p = 1/9$ (where $p$ is the connection probability), (b) an SBM with two $G(N,p)$ ER blocks with $p = 1/9$ and inter-block connection probability $10^{-3}$, (c) an SBM with one $G(N,p)$ ER block with $p = 1/9$, one $G(N,p)$ ER block with $p = 10^{-3}$, and inter-block connection probability $10^{-3}$, (d) a $G(N,k,q)$ WS network in which $q = 1/9$ and each node is adjacent to {$k = 4$} nearest neighbors (where $q$ is the probability of rewiring each edge), and (e) a BA network. 
     All of these networks have $N = 1000$ nodes. In all simulations with weighted networks, we use Hermitian weight matrices (\ie, $W = W^\dagger$). To construct the BA network, we start with a star graph with 1 hub and 2 leaves and iteratively add new nodes until there are $N = 1000$ nodes. Each new node has $m = 2$ edges that connect to existing nodes using linear preferential attachment. The solid orange curves indicate numerical results for binary weight matrices (\ie, for $W = A$). The dotted green curves indicate numerical results for real-valued weight distributions, and the dashed blue curves indicate numerical results for networks with complex-valued weight distributions.
     We distribute the real weights uniformly at random in the interval $[0,4/3)$, and we distribute the complex weights uniformly at random in the first quadrant of the complex plane.
    Each result is a mean of 100 independent instantiations of the indicated random-graph models. 
    For each instantiation, we use the same network structure, but we change the weights (which can be binary, real, or complex).
    The solid gray curves in panel (a) indicate the analytical solution \eqref{eq:semicircle} for $\sigma_2^2 = p(1-p)$, $\sigma_2^2 = 4(4 - 3 p) p/27$, and $\sigma_2^2 = p/2-32 p^2/(9\pi^2)$ [see Eqs.~\eqref{eq:sigma_2_uniform}--\eqref{eq:graph_energy_weighted_complex}].}
    \label{fig:spectrum_1}
\end{figure}
In Fig.~\ref{fig:spectrum_1}, we show the eigenvalue distributions of $W$ for the networks that we studied in Sec.~\ref{sec:graph_energy}. The analytical results in Fig.~\ref{fig:spectrum_1}(a) (see the solid gray curves) are based on a connection between the examined weight matrices and Wigner matrices~\cite{wigner1958distribution}. A Wigner matrix $X_N$ is a real symmetric matrix with entries $x_{ij}$ (with $i,j \in \{1, \ldots, N\}$) that satisfy the following properties~\cite{li2012graph}:
\begin{itemize}
    \item the entries $x_{ij}$ are independent random variables with $x_{ij}=x_{ji}$\,;
    \item the diagonal entries $x_{ii}$ are distributed according to {a distribution $F_1$}, and the off-diagonal entries $x_{ij}$ (with $i\neq j$) are distributed according to {a distribution $F_2$}\,;
    \item the distribution $F_2$ has finite variance $\mathrm{Var}(x_{ij})\equiv \sigma_2^2$ (\ie, $\sigma_2^2 < \infty$).
\end{itemize}
As $N \rightarrow \infty$, the eigenvalue distribution of a normalized Wigner matrix $X_N/N$ converges almost surely to the Wigner semicircle distribution
\begin{equation}     \label{eq:semicircle}
    \phi(x)=\frac{1}{2\pi\sigma_2^2}\sqrt{4 \sigma_2^2-x^2} \, \mathds{1}_{|x|<2\sigma_2}\,,
\end{equation}
where $\mathds{1}_S$ denotes the indicator function on the set $S$.

For the $G(N,p)$ ER random-graph model with binary edge weights, ${\sigma_2^2 = p(1-p)}$, which yields the $\sqrt{p(1-p)}$ term in Eq.~\eqref{eq:ER_energy}.

The weights $w_{ij}$ (with $i \neq j$) of the $G(N,p)$ ER networks with real-valued weight matrices that we studied in Sec.~\ref{sec:graph_energy} are 
\begin{equation}
    w_{ij}=w_{ji}=
	    \begin{cases}
		    x_{ij}\,,\quad& \mathrm{with~probability}~p\\
		    0\,,\quad &\mathrm{with~probability}~1-p\,,
	    \end{cases}
\end{equation}
where $x_{ij} \sim \mathcal{U}[a,b)$ and $\mathcal{U}[a,b)$ denotes the uniform distribution on the interval $[a,b)$. The corresponding variance is
\begin{align} \label{eq:sigma_2_uniform}
\begin{split}
	    \sigma_2^2 &= \frac{p}{b-a} \int_a^b x^2 \,\mathrm{d}x-\frac{p^2(b-a)^2}{4}\\
	&=\frac{1}{12}p\left[4(a^2+a b+b^2)-3(a-b)^2p\right]\,.
\end{split}
\end{align}
For $x_{ij}\sim \mathcal{U}[0,4/3)$ (with $i\neq j$), the energy of almost every $G(N,p)$ ER network with real-valued weights (see Sec.~\ref{sec:graph_energy}) is
\begin{equation}
    E[G(N,p)]=N^{3/2}\left(\frac{8}{3\pi}\sqrt{\frac{4(4 - 3 p) p}{27}}+\smallO(1)\right)\,.
\label{eq:graph_energy_weighted}
\end{equation}
For the $G(N,p)$ ER network with complex-valued weights that we studied in the main manuscript, a similar calculation yields
\begin{equation}
    E[G(N,p)]=N^{3/2}\left(\frac{8}{3\pi}\sqrt{\frac{p}{2}-\frac{32 p^2}{9 \pi^2}}+\smallO(1)\right)\,,
\label{eq:graph_energy_weighted_complex}
\end{equation}
which we obtained using the relations $\mathrm{Var}(Z)=\mathrm{Var}(\Re(Z))+\mathrm{Var}(\Im(Z))$ and $\mathrm{Var}(XY)=(\sigma_X^2+\mu_X^2)(\sigma_Y^2+\mu_Y^2)-\mu_X^2\mu_Y^2$, where $\mu_X$ and $\sigma_X^2$ denote the mean and variance of the random variable $X$. 
Initially, Wigner derived the semicircle law \eqref{eq:semicircle} for real symmetric matrices~\cite{wigner1958distribution}. Subsequently, researchers have examined generalizations to complex-valued Hermitian matrices (see, \eg, Ref.~\cite{bai2005convergence}).

%%%%%

\section{Erd\H{o}s--R\'enyi (ER) graph energy with $p = 1$}
\label{app:ER_graph_energy}
As $p \rightarrow 1$, a binary $G(N,p)$ ER network approaches a complete graph, which has the adjacency matrix
\begin{equation}
    A = u u^\top - I_N\,,
\end{equation}
where $u\in\mathbb{R}^N$ denotes a vector whose entries are all equal to $1$ and $I_N$ denotes the $N \times N$ identity matrix.
We calculate the eigenvalues of $A$ by distinguishing two cases of the eigenvalue equation $Av=\langle u,v\rangle u-v=\lambda v$. For the $N-1$ eigenvectors that are orthogonal to $u$, we have $\langle u,v\rangle=0$. The corresponding eigenvalue (of multiplicity $N-1$) is $-1$. The remaining eigenvector $v=u$ is associated with the eigenvalue $N-1$.

As $p \rightarrow 1$, the energy of a binary $G(N,p)$ ER network approaches the energy of a complete graph, which is
\begin{equation}
   % E[G(N,1)] = 
    \sum_{i = 1}^N |\lambda_i|=2(N - 1)\,.
\end{equation}
%

%%%%%

\section{Absence of strong Perron--Frobenius property in Hermitian two-node networks with complex edge weights with nonzero imaginary parts}

\label{app:strong_perron_frobenius}

Consider a general two-node network with complex edge weights and the Hermitian weight matrix
\begin{equation} \label{eq:2x2_hermitian_weight}
    W=
		\begin{pmatrix}
			a & b e^{\ii\varphi} \\
			b e^{-\ii\varphi} & c
			\end{pmatrix}\,,
\end{equation}
where $a,b,c\in\mathbb{R}_{\geq 0}$ and $\varphi\in[0,2\pi)$. Equation~\eqref{eq:2x2_hermitian_weight} allows self-weights. The largest eigenvalue of $W$ is
\begin{equation}
    \frac{1}{2} \left(a + c \sqrt{(a - c)^2+4 b^2}\right)\,,
\end{equation}
and the corresponding eigenvector is
\begin{equation}
    \left(\frac{e^{\ii \varphi}(a-c + \sqrt{(a - c)^2+4 b^2})}{2 b}, 1\right)^\top\,.
\label{eq:2x2_ev}
\end{equation}
The imaginary part of the off-diagonal components of $W$ is $0$ for $\varphi = 0$ and $\varphi = \pi$. However, for the eigenvector \eqref{eq:2x2_ev} to be positive, the phase $\varphi$ must be either $0$ or $\pi$. Therefore, the imaginary part of the off-diagonal entries of $W$ [see Eq.~\eqref{eq:2x2_hermitian_weight}] is $0$. Consequently, it is not possible for a two-node network with a Hermitian weight matrix to have off-diagonal matrix entries with nonzero imaginary part and also satisfy the strong Perron--Frobenius property~\cite{noutsos2012perron}.
%
%%%%%%%

\section{Quantum random-walk betweenness and closeness centralities}
\label{app:betweenness_closeness}
In the main manuscript, we used the infinite-time mean \eqref{eq:quant_occ} of a CTQW [see Eq.~\eqref{eq:CTQW}] to define the occupation centralities of the nodes of a network with a Hermitian weight matrix $W$. To characterize betweenness and closeness centralities of the nodes of a network with complex weights, we first define the absorbing quantum random-walk Hamiltonian
\begin{align} \label{eq:H_absorbing}
    [(H_{\ell})^{\rm a}_{\rm q}]_{ij} =
		\begin{cases}
			(H_{\rm q})_{ij}\,,\quad &\text{if $j \neq \ell$}\\
			0\,,\quad &\text{if $j = \ell$}
		\end{cases}
\end{align}
with absorbing node $\ell$. The basic idea that underlies the use of an absorbing Hamiltonian is that we wish to track the number of times that a quantum random walker traverses a node if its final destination is $\ell$~\cite{sole2016random,boettcher2021classical}. 
Taking a mean over all absorbing nodes yields a measure of random-walk betweenness centrality. Note that $(H_{\ell})^{\rm a}_{\rm q}$ is a non-Hermitian operator. To evaluate its infinite-time mean \eqref{eq:quant_occ}, we treat the upper triangular part of $(H_{\ell})^{\rm a}_{\rm q}$ as equal to the conjugate transpose of the lower triangular part. To do so, we use eigenvalue-problem solvers such as \texttt{scipy.linalg.eigh} ({\sc scipy} version 1.9.1) and \texttt{numpy.linalg.eigh} ({\sc numpy} version 1.23), which treat non-Hermitian matrices as Hermitian matrices. We denote the corresponding Hermitian version of $(H_{\ell})^{\rm a}_{\rm q}$ by $(\widetilde{H}_{\ell})^{\rm a}_{\rm q}$. 

The quantum random-walk betweenness centrality of node $j$ is
\begin{widetext}
\begin{equation} \label{eq:qw_betweenness}
    \tau_{j} = \lim_{s \rightarrow 0}\frac{1}{N(N-1)} \sum_{\ell} \sum_{m,n} \frac{\braket{e_m^{(\ell)}|\psi(0)}  \braket{\psi(0)|e_n^{(\ell)}}}{s+\ii(\lambda_m^{(\ell)}-\lambda_n^{(\ell)})}\braket{j|e_m^{(\ell)}} \braket{e_n^{(\ell)}|j}\,,
\end{equation}
\end{widetext}
where $\ket{\psi(0)}=(1,\ldots,1)^\top/\sqrt{N}$ and $e_m^{(\ell)}$ and $\lambda_m^{(\ell)}$, respectively, are the orthonormal eigenvectors and corresponding eigenvalues of the Hamiltonian $(\widetilde{H}_{\ell})^{\rm a}_{\rm q}$~\cite{boettcher2021classical}. That is,
\begin{equation}
   (\widetilde{H}_{\ell})^{\rm a}_{\rm q} e_m^{(\ell)} = \lambda_m^{(\ell)} e_m^{(\ell)}\,,\quad \braket{e_m^{(\ell)}|e_n^{(\ell)}}=\delta_{mn}\,.
\end{equation}
We use the regularization parameter $s$ in Eq.~\eqref{eq:qw_betweenness} to prevent the denominator from equaling $0$ when $\lambda_m^{(\ell)}=\lambda_n^{(\ell)}$. In all of our numerical experiments, we set $s = 10^{-1}$.

The quantum random-walk closeness centrality of node $\ell$ is equal to the inverse of the mean first-passage time $h_{\ell}$ of a walker that starts at any node and stops after reaching node $\ell$. The mean first-passage time is 
\begin{align}
    h_{\ell} =\frac{1}{N-1} \sum_{i,j} (\tau_{\ell})_{ij}+ \frac{1}{N} \pi_{\ell}^{-1}\,,
    \label{eq:closeness}
\end{align}
where $(\tau_{\ell})_{ij}$ denotes the expected number of times that a random walker that starts at node $i$ with final destination $\ell$ traverses node $j$ at any time. As in Eq.~\eqref{eq:qw_betweenness}, we do not sum over all initial nodes $i$; instead, we consider a uniform initial walker state $\ket{\psi(0)}=(1,\ldots,1)^\top/\sqrt{N}$ and compute $\sum_j (\tau_{\ell})_{j}$ in terms of Eq.~\eqref{eq:qw_betweenness} by replacing the sum over $\ell$ with a sum over $j$.
The quantity $\pi_{\ell}^{-1}$ is the inverse of the occupation probability \eqref{eq:quant_occ} (\ie, the mean return time). 

For more information about classical and quantum random-walk centrality measures and their generalizations to multilayer networks, see Refs.~\cite{sole2016random,boettcher2021classical}.

%%%%

%
\bibliography{refs-rev-v03.bib}
\end{document}